\documentclass{aa}

\usepackage{natbib}
\bibpunct{(}{)}{;}{a}{}{,} 
\usepackage{graphicx}   
\usepackage[varg]{txfonts}
\usepackage{hyperref}   

\hypersetup{
    colorlinks=true,       
    linkcolor=blue,          
    citecolor=blue,        
    filecolor=blue,      
    urlcolor=blue           
}

\def\ga{\,\,\raise0.14em\hbox{$>$}\kern-0.76em\lower0.28em\hbox{$\sim$}\,\,}
\def\la{\,\,\raise0.14em\hbox{$<$}\kern-0.76em\lower0.28em\hbox{$\sim$}\,\,}

\def\cm3{cm$^{-3}$}

\def\chem#1#2{$\mathrm{^{#2}\kern-0.8pt#1}$}
\def\reac#1#2#3#4#5#6{$\mathrm{\, ^{#2}\kern-0.8pt{#1}\, ({#3}\, ,{#4})\, {}^{#6}\kern-0.8pt{#5}\, }$}

\def\be{\begin{equation}} 
\def\ee{\end{equation}}
\def\beqy{\begin{eqnarray}}
\def\eeqy{\end{eqnarray}}
\def\bmlet{\begin{mathletters}}
\def\emlet{\end{mathletters}}

\begin{document}

\title{The p-process in exploding rotating massive stars}

\author{   
A. Choplin\inst{1}
\and S. Goriely\inst{1}
\and R. Hirschi\inst{2,5,6}
\and N. Tominaga\inst{3,4,5,7}
\and G. Meynet\inst{8}
}

        \institute{Institut d'Astronomie et d'Astrophysique, Universit\'e Libre de Bruxelles,  CP 226, B-1050 Brussels, Belgium
        \and Astrophysics Group, Lennard-Jones Labs 2.09, Keele University, ST5 5BG, Staffordshire, UK
        \and National Astronomical Observatory of Japan, National Institutes of Natural Sciences, 2-21-1 Osawa, Mitaka, Tokyo 181-8588, Japan
        \and Department of Astronomical Science, School of Physical Sciences, The Graduate University of Advanced Studies (SOKENDAI), 2-21-1 Osawa, Mitaka, Tokyo 181-8588, Japan
        \and Kavli Institute for the Physics and Mathematics of the Universe (WPI), The University of Tokyo, 5-1-5 Kashiwanoha, Kashiwa, Chiba 277-8583, Japan
        \and UK Network for Bridging the Disciplines of Galactic Chemical Evolution (BRIDGCE)
        \and Department of Physics, Faculty of Science and Engineering, Konan University, 8-9-1 Okamoto, Kobe, Hyogo 658-8501, Japan
        \and Geneva Observatory, University of Geneva, Chemin Pegasi 51, CH-1290 Sauverny, Switzerland\\
                }

\date{Received --; accepted --}

\abstract
{
The p-process nucleosynthesis can explain proton-rich isotopes that  are heavier than iron, which are observed in the Solar System, but discrepancies still persist (e.g. for the Mo and Ru p-isotopes), and some important questions concerning the astrophysical site(s) of the p-process remain unanswered.
}
{
We investigate how the p-process operates in exploding rotating massive stars that have experienced an enhanced s-process nucleosynthesis during their life through rotational mixing. 
}
{
With the Geneva stellar evolution code, we computed 25~$M_{\odot}$ stellar models at a metallicity of $Z=10^{-3}$ with different initial rotation velocities and rates for the still largely uncertain $^{17}$O($\alpha$,$\gamma$)$^{21}$Ne reaction. 
The nucleosynthesis calculation, followed with a network of 737 isotopes, was coupled to stellar evolution, and  
the p-process nucleosynthesis was calculated in post-processing during both the final evolutionary stages and spherical explosions of various energies. The explosions were modelled with a relativistic hydrodynamical code.
}
{
In our models, the p-nuclides are mainly synthesized during the explosion, but not much during the ultimate hydrostatic burning stages. 
The p-process yields mostly depend on the initial number of trans-iron seeds, which in turn depend on the initial rotation rate.
We found that the impact of rotation on the p-process is comparable to the impact of rotation on the s-process. 
From no to fast rotation, the s-process yields of nuclides with mass number $A<140$ increase by $3-4$ dex, and so do the p-process yields. Fast rotation with a lower $^{17}$O($\alpha,\gamma$) rate significantly produces s- and p-nuclides with $A\geq140$.
The dependence of the p-process yields on the explosion energy is very weak.
}
{
Our results suggest that the contribution of core-collapse supernovae from massive stars to the solar (and Galactic) p-nuclei has been underestimated in the past, and more specifically, that the contribution from massive stars with sub-solar metallicities may even dominate. A more detailed study including stellar models with a wide range of masses and metallicities remains to be performed, together with a quantitative analysis that is based on the chemical evolution of the Galaxy.
}


\titlerunning{The p-process in exploding rotating massive stars}

 \authorrunning{Choplin et al.}

\maketitle


\section{Introduction}
\label{sect:intro}

Despite tremendous progress during the past decades, the origin of the trans-iron chemical elements is still debated and not yet fully understood \citep[e.g.][]{arnould20}. 
The slow (s) and rapid (r) neutron capture processes are the two main processes that have each forged about the half of the trans-iron nuclides. 
The s-process \citep[e.g. the review of][]{kappeler11} operates during the late life of asymptotic giant branch (AGB) stars \citep[main s-process; e.g.][]{gallino98,herwig05,cristallo11,karakas14} and during the core helium-burning and shell carbon-burning stages of massive stars \citep[weak s-process; e.g.][]{langer89, prantzos90,raiteri91b,the07}. The r-process is associated with explosive events such as neutron star mergers \citep[e.g.][]{arnould07,goriely11b,wanajo14,just15}, magnetorotational supernovae \citep{winteler12,nishimura15}, or collapsars \citep{siegel19}. 
At neutron densities in between the s- and r-processes, the existence of an intermediate (i) neutron capture process \citep[first named by][]{cowan77} is expected. 
Its astrophysical site(s) is (are) actively debated \citep[see Sect.~1 of][for a list of possible sites]{choplin21}.
Other nuclear processes also include short but possibly intense neutron bursts taking place in the helium shell of exploding massive stars \citep{blake76, thielemann79, meyer04, choplin20}. This process is thought to cause the anomalous isotopic signatures found in meteorites \citep{meyer00,pignatari15,pignatari18}, and possibly the abundances of some metal-poor r/s-stars \citep{choplin20}.

\begin{figure*}
\includegraphics[width=\columnwidth]{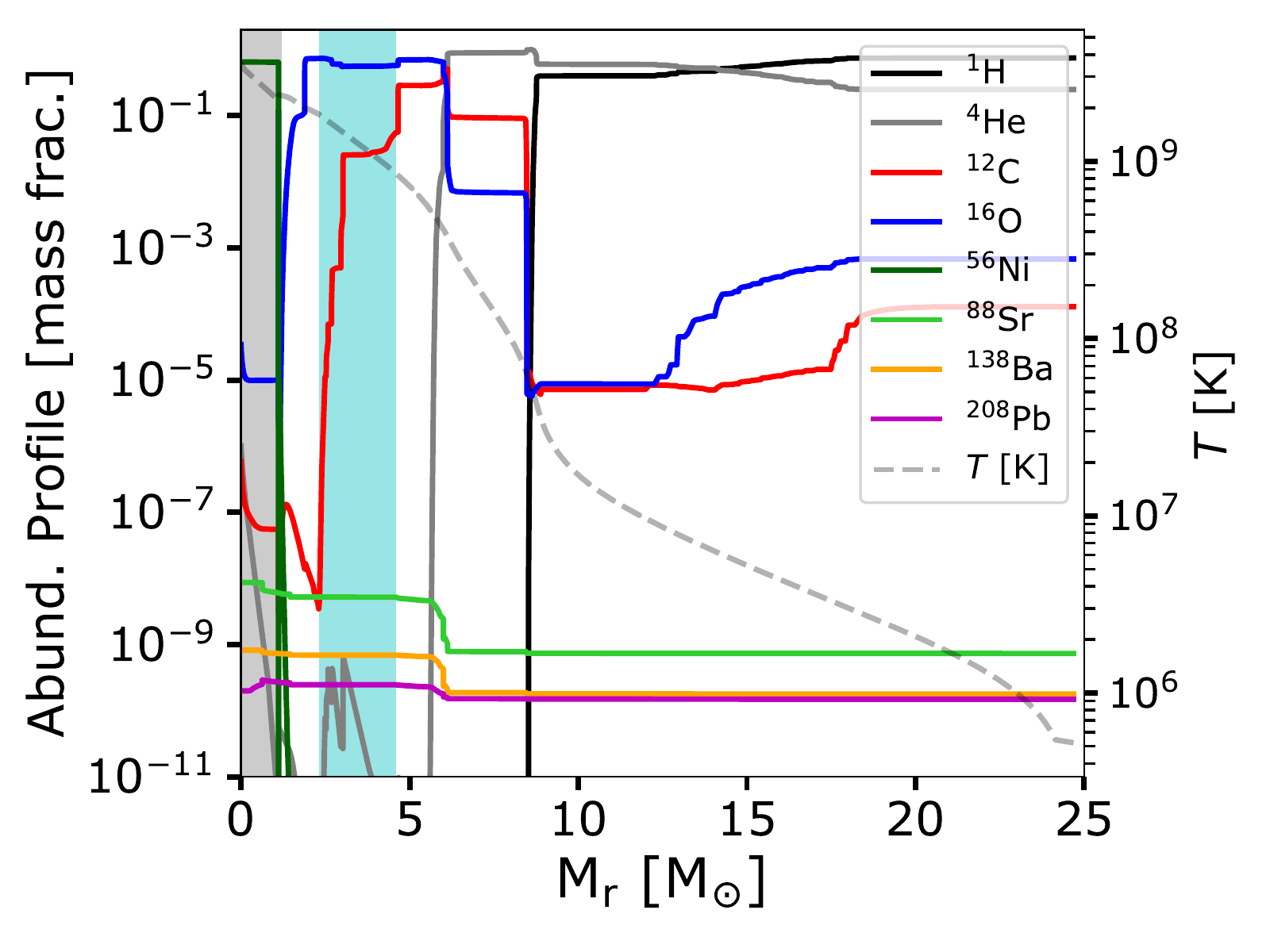}
\includegraphics[width=\columnwidth]{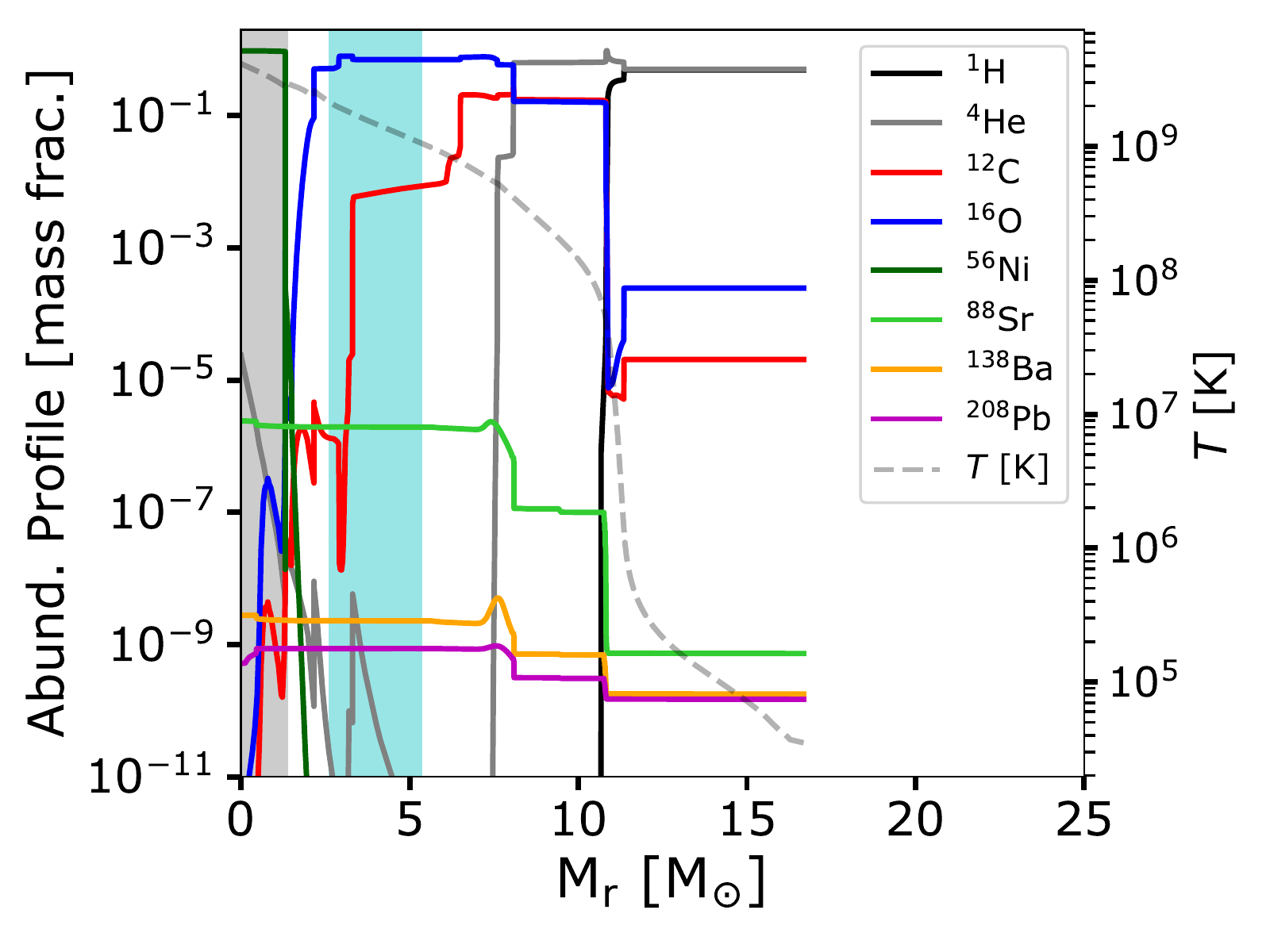}
\caption{Pre-supernova abundance profiles for the M25Zm3V0 (non-rotating, left panel) and M25Zm3V4 (rotating, $\upsilon_{\rm ini}/ \upsilon_{\rm crit} = 0.4$, right panel) models computed with the rate of $^{17}$O($\alpha$,$\gamma$)$^{21}$Ne from \cite{best13}. 
The abundances of trans-iron elements are not explicitly followed from core O-burning ignition (see text for details). 
The dashed grey line shows the pre-supernova temperature profile, the scale of which is indicated on the right axis.
The shaded grey area represents the remnant. 
The cyan area represents the zone in which the maximum temperature during a spherical $10^{51}$~erg explosion ranges between 1.8 and 3.7~GK (i.e. the zone in which the explosive p-process takes place). 
}
\label{fig_ab1}
\end{figure*}

Finally, the p-process \citep{arnould76,woosley78,meyer94,arnould03}  is thought to take place in the hydrostatic oxygen- and neon-burning shells of massive stars \citep{arnould76,rauscher02}, during core-collapse supernovae \citep[CCSNe;][]{rayet95}, or during type Ia supernovae \citep{travaglio15}. 
The p-process occurs through combinations of ($\gamma$,$n$), ($\gamma$,$p$), and ($\gamma$,$\alpha$) reactions affecting pre-existing s- or r-nuclides. 
The p-process provides a possible explanation for the neutron-deficient nuclides in Solar System abundances \citep[e.g.][]{arnould03,lugaro16}. 
Some discrepancies remain, however. 
In particular,  $^{92,94}$Mo and $^{96,98}$Ru p-isotopes are systematically underproduced \citep[][their Fig.~22]{rayet95,arnould03,arnould20}. 
This fact motivates the search for alternative or additional ways to produce these nuclides. In particular, the so-called pn process, i.e. a proton-poor neutron-boosted rp-process, which is encountered during He detonation, has been suggested as a promising nucleosynthesis source \citep{goriely05b}. Such an object is made of a carbon-oxygen white dwarf with a sub-Chandrasekhar mass ($M < 1.4 M_\odot$) that accumulates a He-rich layer at its surface. An alternative site proposed to explain the origin of the Mo and Ru p-nuclei is the p-rich neutrino-driven wind  in CCSNe, where antineutrino absorptions in the proton-rich environment produce neutrons that are immediately captured by neutron-deficient nuclei \citep[the $\nu p$-process;][]{frohlich06,ghosh21}.
Recent results of \cite{bliss18} suggest that these proton-rich winds can make dominant contributions to the solar abundance of $^{98}$Ru, but that additional astrophysical sources are likely required to account for $^{92,94}$Mo and $^{96}$Ru. The proton richness of the wind also remains a highly debated question. 

\cite{costa00} have shown that an increase in $^{22}$Ne($\alpha$,$n$)$^{25}$Mg reaction during stellar evolution leads to a pre-supernova seed distribution that can ultimately provide enough $^{92,94}$Mo and $^{96,98}$Ru p-isotopes during the explosion.
Interestingly, the $^{22}$Ne($\alpha$,$n$)$^{25}$Mg reaction can be naturally boosted in rotating massive stars through additional production of $^{22}$Ne during core-helium burning \citep{meynet06,hirschi07,hirschi08}. The efficiency of the weak s-process is thus significantly boosted if the massive star is rotating \citep{pignatari08,frischknecht16,choplin18,limongi18,banerjee19}. 
In rotating massive stars, the pre-supernova distribution of trans-iron elements is therefore different from that in non-rotating massive stars. 
This will impact the p-process that can take place during the explosion of such stars, and which largely depends on the  trans-iron seeds before the explosion.

In this paper we explore how rotation impacts the p-process nucleosynthesis in exploding rotating massive stars and whether these stars can account for the abundances of solar p-nuclides. Section~\ref{sect_mod} presents the input physics relevant to the present work, Sect.~\ref{sect_res} discusses the results obtained in terms of s-process during the hydrostatic burning stages and of p-process during both the hydrostatic and explosive evolutions. 
Section~\ref{sect_pcontrib} investigates the contribution of massive rotating stars to the Galactic p-nuclei. 
Conclusions are given in Sect.~\ref{sect_concl}.

\begin{figure}
\includegraphics[scale=0.52]{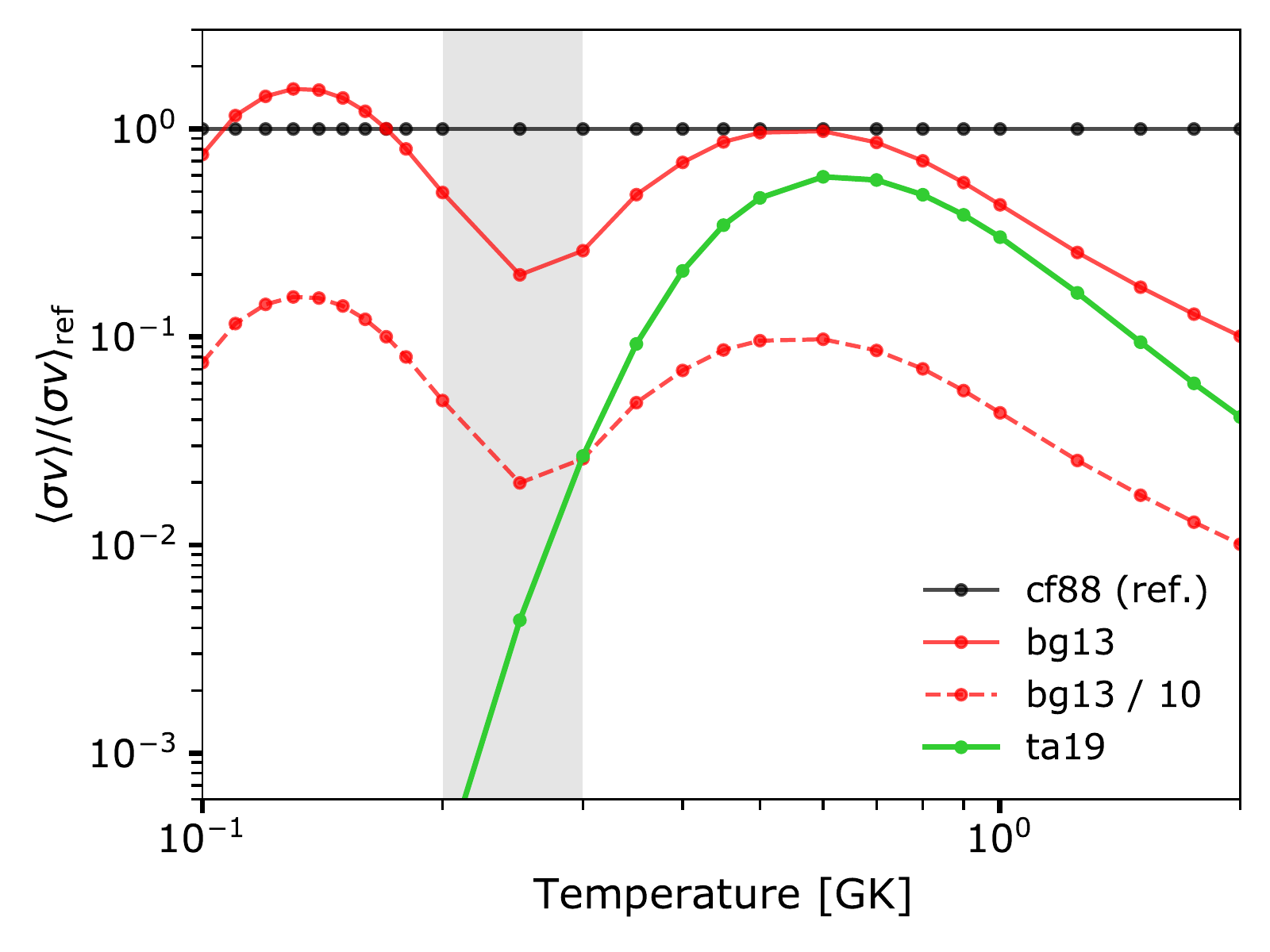}
\caption{
Ratio of $^{17}$O($\alpha$,$\gamma$)$^{21}$Ne reaction rates to the rate of \citet[][cf88 label]{caughlan88}. 
The red curve (label bg13) is the recommended rate of \cite{best13}. 
The green curve (label ta19) is the experimental lower limit from \cite{taggart19}. 
The shaded area indicates the approximate range of temperatures in the helium-burning core of massive stars.
}
\label{fig_o17}
\end{figure}

\begin{table}
\scriptsize{
\caption{Characteristics of our 25~$M_{\odot}$,~$Z=10^{-3}$ stellar models at the end of their evolution. 
$M_{\rm CO}$ (column 6) is the mass of the carbon-oxygen core, defined where the mass fraction of $^{4}$He has dropped below $10^{-2}$. 
$M_{\rm Ni}$ (column 7) is the mass of the Ni core, defined when the mass fraction of $^{56}$Ni drops below $10^{-2}$ (the models without a value for $M_{\rm Ni}$ were not computed until the very end of the evolution; see text for details). 
\label{table:1}
}
\begin{center}
\resizebox{8.5cm}{!} {
\begin{tabular}{lcccc} 
\hline
Model label   &   $\upsilon_{\rm ini}/ \upsilon_{\rm crit} $    & $^{17}$O($\alpha$,$\gamma$)$^{21}$Ne &  $M_{\rm CO}$  &  $M_{\rm Ni}$          \\
  &   &    &  [$M_{\odot}$]  &  [$M_{\odot}$]        \\
\hline
M25Zm3V0    &  0   &  \cite{best13}          &  5.90   &   1.11       \\
M25Zm3V4   &  0.4   & \cite{best13}        &   7.61  &   1.31     \\
M25Zm3V7   &  0.7  &   \cite{best13}       &  7.56    &   $-$    \\
\hline
M25Zm3V0B &    0  & \cite{best13} / 10  &   5.88   &      $-$      \\
M25Zm3V4B &    0.4  & \cite{best13} / 10  &   7.64   &   $-$         \\
M25Zm3V7B &    0.7  & \cite{best13} / 10  &  7.62    &   $-$      \\
\hline
M25Zm3V0C  &  0  & \cite{taggart19}  &  5.88    &      $-$      \\
M25Zm3V4C  &  0.4  & \cite{taggart19}  &  7.65    &   $-$         \\
M25Zm3V7C  &  0.7  & \cite{taggart19}  &  7.57    &   $-$      \\
\hline
\end{tabular}
}
\end{center}
}
\end{table}


\section{Input physics}
\label{sect_mod}

\subsection{Stellar evolution models}

We started from two $25$~$M_{\odot}$ models at a metallicity of $Z=10^{-3}$ in mass fraction, which were computed with the Geneva stellar evolution code \citep{eggenberger08} and published in \citet[][models labelled 25S0 and 25S4 in their Table~1]{choplin18}. 
The first model was non-rotating, and the second model was rotating with\footnote{The critical velocity $\upsilon_{\rm crit}$ is reached when gravitational acceleration is counterbalanced by centrifugal force. In the Roche approximation, it is expressed as $\upsilon_{\rm crit} = \sqrt{\frac{2}{3}\frac{GM}{R_{\rm p,c}}}$ , with $R_{\rm p,c}$ the polar radius at the critical limit.} $\upsilon_{\rm ini}/ \upsilon_{\rm crit} =$ 0.4. 
All details of the input physics can be found in \cite{choplin18}. We recall the main physical ingredients here.

During stellar evolution, a nuclear reaction network of 737 isotopes (from hydrogen to polonium), coupled to the structure equations, was used. 
Opacity tables were computed with the OPAL tool\footnote{https://opalopacity.llnl.gov/} and complemented at low
temperature by opacities from \cite{ferguson05}.
Radiative mass-loss rates were taken from \cite{vink01} if log $T_{\rm eff} \ge 3.9$ and from \cite{jager88} otherwise.
For convection, the Schwarzschild criterion was used. 
During the H- and He-burning phases, the size
of the convective core is extended by $d_{\rm over} = \alpha H_P$ , with $H_P$ the pressure scale height estimated at the Schwarzschild boundary, and $\alpha = 0.1$. Rotation was treated according to the shellular scheme \citep{zahn92, chaboyer92,maeder98}. The diffusion coefficients for horizontal and vertical shear were taken from \cite{zahn92} and \cite{talon97}, respectively.
The rates of $^{22}$Ne($\alpha,n$)$^{25}$Mg and $^{22}$Ne($\alpha,\gamma$)$^{26}$Mg were taken from \cite{longland12}.
The rates of $^{17}$O($\alpha,n$)$^{20}$Ne and $^{17}$O($\alpha,\gamma$)$^{21}$Ne were taken from \cite{best13} if not stated otherwise. 

 \cite{choplin18} evolved the stellar models up to core O-burning ignition with the 737 isotopes network. 
At core O-burning ignition, most of the stellar layers ejected at the time of the supernova have reached their final state and will not be impacted during O- and Si-burning phases. 
However, the innermost layers (below a mass coordinate of about $M_r = 3$~$M_{\odot}$ for the non-rotating 25~$M_{\odot}$ model; Fig.~\ref{fig_ab1}, left panel)  are impacted during these short ultimate stages.
The p-process occurs in relatively deep layers into the star, and for the present work, it is therefore preferable to go further in the evolution to have a more reliable pre-supernova structure of the inner layers. 
We thus continued the evolution until the end of the core Si-burning stage. 
During these ultimate stages, the effect of rotation was switched off. 
This was shown to be a good approximation because the evolutionary timescale is far shorter than the rotational mixing timescale \citep[][their  Fig.~2]{choplin17a}.
We also used the standard minimum network of the Geneva code \citep[e.g.][]{hirschi07, ekstrom12} instead of the full 737 isotope network. This minimum network ensures a proper description of the energetics and keeps track of the main isotopes. The pre-supernova abundance profiles of some specific nuclei are illustrated in Fig.~\ref{fig_ab1}.

\begin{figure*}[t]
\includegraphics[width=\columnwidth]{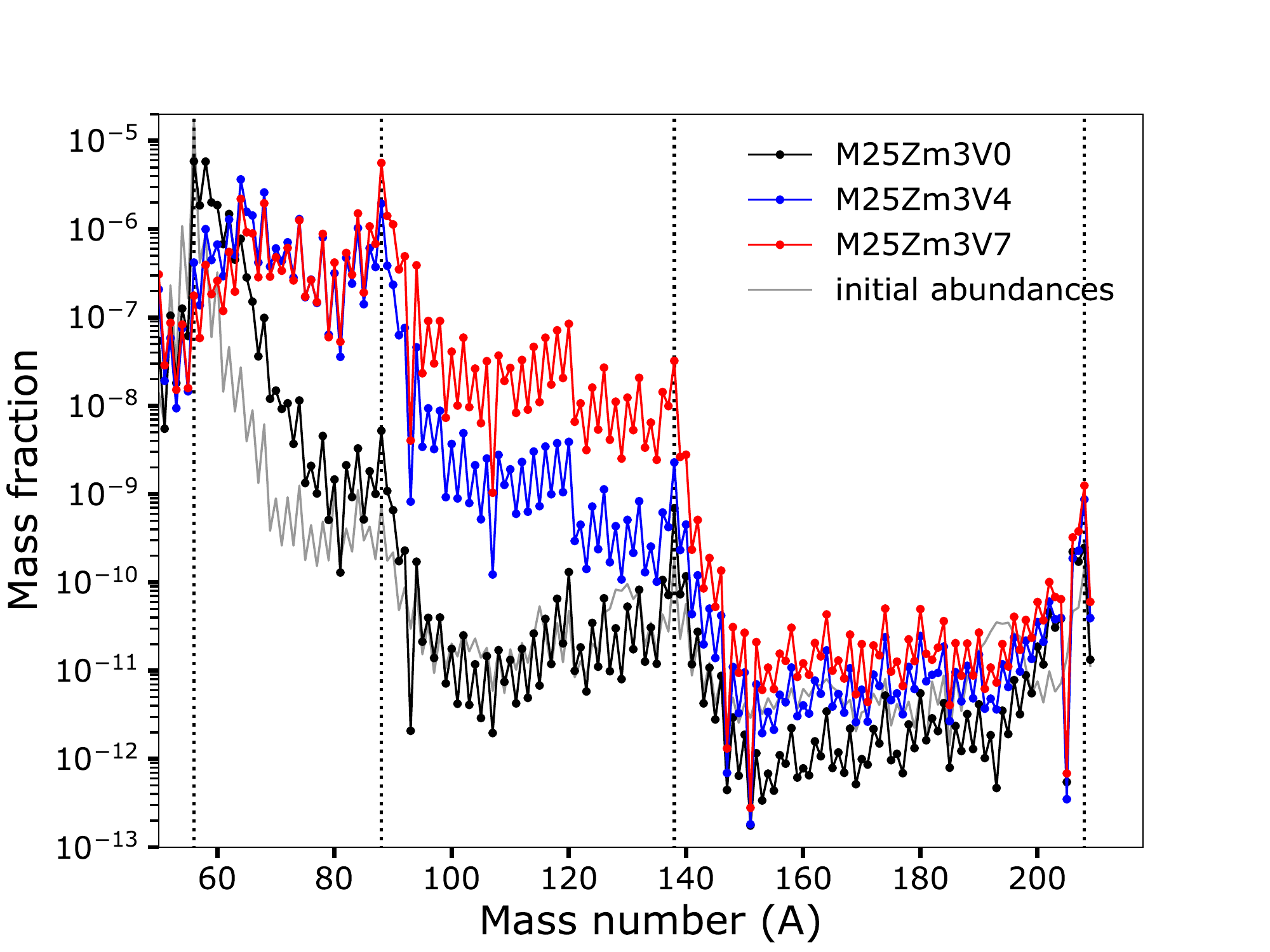}
\includegraphics[width=\columnwidth]{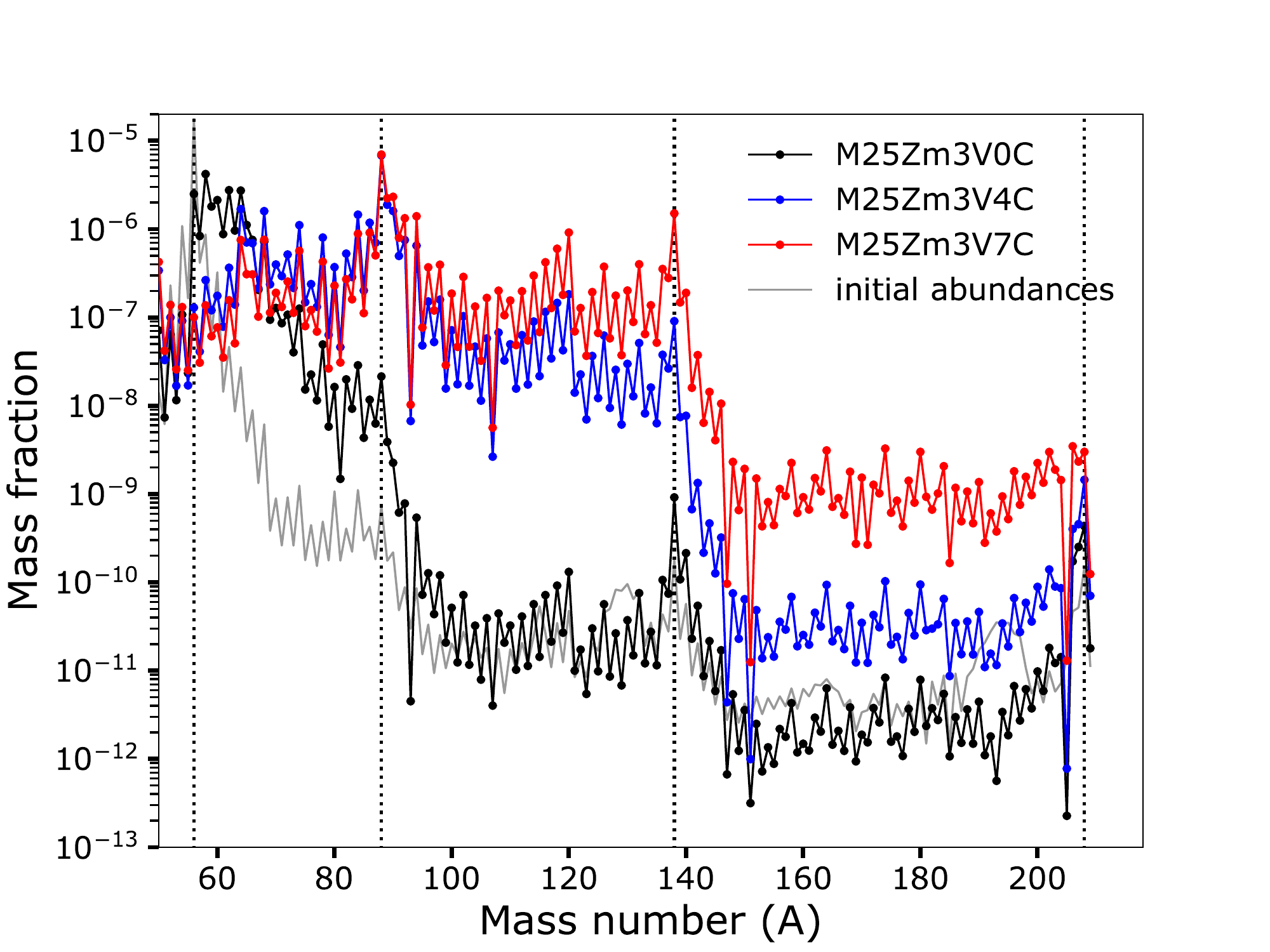}
\caption{
Pre-supernova mass fractions of elements heavier than iron as a function of the mass number $A$ at a mass coordinate of 4~$M_{\odot}$ for the models computed with the rate of $^{17}$O($\alpha$,$\gamma$)$^{21}$Ne from \citet[][left panel]{best13} and \citet[][right panel]{taggart19}. The four vertical dashed lines show the location of the s-nuclides $^{56}$Fe, $^{88}$Sr, $^{138}$Ba, and $^{208}$Pb. 
}
\label{fig_ab3}
\end{figure*}

In addition to the two reference $25$~$M_{\odot}$ models (with $\upsilon_{\rm ini}/ \upsilon_{\rm crit} = 0$ and $0.4$), we investigated the case of  a model with the same initial properties that was a faster rotator, however, with $\upsilon_{\rm ini}/ \upsilon_{\rm crit} = 0.7$.
Unlike the first two models, the evolution of this model was not followed after core O-burning ignition. We adopted the same explosion model as for the $\upsilon_{\rm ini}/ \upsilon_{\rm crit} = 0.4$ model, but with the important difference that the initial abundances of trans-iron elements prior to the explosion were not the same. 
This means that the same mass cut\footnote{At the time of the explosion, the mass cut is the mass coordinate that delimits the part of the star that is expelled from the part that is locked into the remnant.} as for the $\upsilon_{\rm ini}/ \upsilon_{\rm crit} = 0.4$ model was adopted and the temperature and density histories of the stellar layers during the explosion were the same as for the $\upsilon_{\rm ini}/ \upsilon_{\rm crit} = 0.4$ model (cf. Sect.~\ref{sect:explomod} for more details of the explosive modelling).
This is a good approximation because the $\upsilon_{\rm ini}/ \upsilon_{\rm crit} = 0.4$ and 0.7 models have very similar characteristics at core oxygen-burning ignition (in particular, similar temperature and density profiles, and similar CO-core masses). 
The only important difference between the $\upsilon_{\rm ini}/ \upsilon_{\rm crit} = 0.4$ and 0.7 models is the abundance of trans-iron elements prior to the explosion (Fig.~\ref{fig_ab3} and Sect.~\ref{sect_st_mod}).

For each of our three model stars, we also considered two alternative models for which the rate of the $^{17}$O($\alpha$,$\gamma$)$^{21}$Ne reaction was changed, as illustrated in Fig.~\ref{fig_o17}.
In one case (models B), we used the rate from \cite{best13} divided by 10 instead of the original rate from \cite{best13}.  
In a second case (models C) we adopted the lower limit from \cite{taggart19} instead of \cite{best13}.
Although theoretical and experimental works were carried out to study this rate \citep{descouvemont93,best11,best13,taggart19}, significant uncertainties still remain in the temperature range of interest for the s-process. 
The uncertainty on this rate was shown to dramatically affect the s-process efficiency in rotating massive stars \citep[e.g.][]{taggart19}.
A low $^{17}$O($\alpha$,$\gamma$)$^{21}$Ne rate enhances the s-process efficiency because in this case, the competing  $^{17}$O($\alpha$,n)$^{20}$Ne reaction becomes dominant and recycles  neutrons \citep[more details in Sect.~\ref{sect_st_mod}, also in Sect.~3.5 of][]{choplin18}. 
Only a theoretical estimate of $^{17}$O($\alpha$,$\gamma$)$^{21}$Ne is available at low temperature \citep[][black and red lines in Fig.~\ref{fig_o17}]{caughlan88,best13}. \cite{taggart19} constrained its lower limit experimentally. In the temperature range of interest for the s-process, it is 10 to 1000 times lower than the recommended rate of \citet[][green line in Fig.~\ref{fig_o17}]{best13}. 

\begin{figure}[t]
\includegraphics[width=\columnwidth]{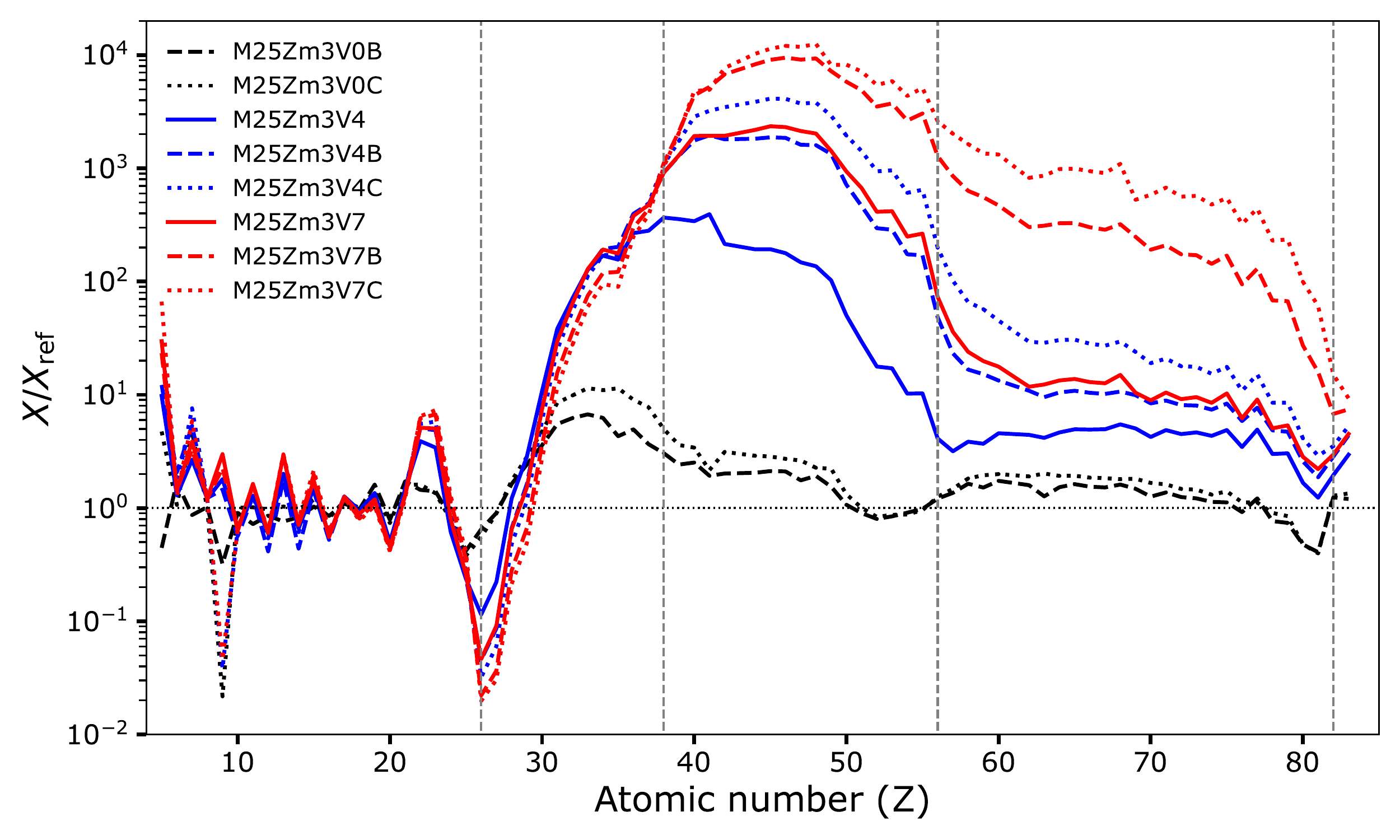}
\caption{
Elemental mass fractions of all models normalized by the mass fractions of the reference M25Zm3V0 model. Like in Fig.~\ref{fig_ab3}, the abundances are shown at a mass coordinate of 4~$M_{\odot}$. The four vertical dashed lines show the location of Fe ($Z=26$), Sr ($Z=38$), Ba ($Z=56$), and Pb ($Z=82$). 
}
\label{fig_ab4}
\end{figure}

As mentioned above, in comparison with the standard models, the sets of models computed with the \cite{best13} rate divided by 10 or the \cite{taggart19} rate experience a more efficient s-process during the evolution because the  $^{17}$O($\alpha$,$\gamma$)$^{21}$Ne rate is lower (Fig.~\ref{fig_o17}).
Like for the $\upsilon_{\rm ini}/ \upsilon_{\rm crit} = 0.7$ model with the rate of \cite{best13}, the evolution after core O-burning ignition is not followed for these six additional models. 
We rely on the explosion of the corresponding models computed with the \cite{best13} rate.
Here again, this is a good approximation because for a given $\upsilon_{\rm ini}/ \upsilon_{\rm crit} $ ratio, the models with different  $^{17}$O($\alpha$,$\gamma$)$^{21}$Ne rates behaves similarly. The pre-supernova abundance distributions for the different models are compared in Figs.~\ref{fig_ab3}-\ref{fig_ab4} and are further discussed in Sect.~\ref{sect_st_mod}. 
Table~\ref{table:1} summarizes the main characteristics of the nine models considered in this work.

\subsection{Explosion models}
\label{sect:explomod}

Two common and simple approximations for simulating the explosion of massive stars are the piston model \citep[e.g.][]{woosley95, limongi03, heger10} and the thermal bomb \citep[e.g.][]{thielemann96,tominaga07a}. 
The piston model usually imitates the collapse and bounce of the inner boundary. The radius of inner boundary is first reduced and then increased. 
In the second method, the energy is deposited at the mass cut of the exploding star in the form of internal energy. 
More recent methods are the so-called PUSH \cite[e.g.][]{perego15,ebinger19,curtis19} or P-HOTB \cite[e.g.][]{sukhbold16,ertl20} approaches, in which the mass cut and explosion energy can emerge from the simulation instead of being assumed.

In this work, the explosion energy is deposited as kinetic energy at the chosen mass cut. We used the relativistic hydrodynamical code from \cite{tominaga07a} and \cite{tominaga09}. 
Although this code allows the computation of two dimensional explosions, we considered one-dimensional spherical explosions.
A code like this was already used in \cite{choplin20} to investigate the effect of a jet-like explosion hitting the helium-burning shell of rotating models computed with the Geneva code.

In the present work, a spherical explosion was triggered in the non-rotating and the $\upsilon_{\rm ini}/ \upsilon_{\rm crit} = 0.4$ models with the \cite{best13} rate (M25Zm3V0 and M25Zm3V4 in Table~\ref{table:1}). 
The total energy deposited was $E_{\rm tot} = 10^{51}$~erg in the standard case. 
Different explosion energies are considered in Sect.~\ref{sect_exp_ene}. 
The energy was deposited at a mass coordinate of 1.2 and 1.4~$M_{\odot}$ for the non-rotating and rotating model, respectively. 
These mass coordinates correspond to the top of the $^{56}$Ni core (Fig.~\ref{fig_ab1}). 
The hydrodynamics was followed for $500$~seconds. 
The temperature and density histories are recorded by mass particles representing Lagrangian mass elements of the stellar mantle. 

\subsection{p-process nucleosynthesis}

During the last hydrostatic burning phases of massive stars and in the most inner layers, the trans-iron elements progressively photodisintegrate through a combination of ($\gamma,n$), ($\gamma,\alpha$), and ($\gamma,p$) reactions. 
It was shown that some p-nuclides could form before the supernova explosion in the oxygen-burning shell and that this process is very sensitive to initial mass or to the convection model \citep{arnould76,rauscher02}.
The production of p-nuclides during the explosion takes place at higher radii than during the hydrostatic evolution (in the layers that are sufficiently heated by the supernova wave).
We considered both production channels (during the evolution and during the explosion). The p-process nucleosynthesis was treated in post-processing calculations during the late hydrostatic evolution (oxygen-burning onwards) and during the explosion. 

The nucleosynthesis was calculated with a code that was especially designed to follow all reactions of relevance during the p-process nucleosynthesis \citep{arnould03}. Changes in composition were followed by a full network calculation including all 2200 species from protons up to $Z=84$ that lie between the proton-drip line and the neutron-rich region that may be populated. All neutron and charged-particle fusion reactions as well as their reverse reactions on elements up to Po isotopes were included.  The reaction rates on light species were taken from the NETGEN library, which includes all the latest compilations of experimentally determined reaction rates \citep{Xu13}. 
Experimentally unknown reaction rates were estimated with the TALYS code \citep{Koning12,Goriely08} on the basis of the Skyrme Hartree-Fock-Bogolyubov (HFB) nuclear mass model, 
HFB-24 \citep{Goriely13a}, when they were not available experimentally. In addition to these reactions, electron captures and $\beta$-decays were also included. The corresponding rates were taken from experimental data \citep{Kondev21} when available, and from the gross theory \citep{Tachibana90}, otherwise.

p-process nuclei are produced exclusively in layers heated at temperatures ranging between typically 1.8 and $3.5~\times~10^9$~K \citep{rayet90,rayet95,arnould03}. Below $1.8$~GK, the heavy seeds do not efficiently photodisintegrate during the evolutionary timescale. Above $3.5$~GK, all seeds are photodisintegrated into iron-peak elements.
For this reason, only layers with $1.8 \la T_{\rm max} \la 3.7$~GK (where $T_{\rm max}$ is the maximum temperature reached during the hydrostatic evolution and/or explosion) were post-processed in the present study. These layers are referred to as p-process layers (PPLs). 

In this work, the production of a p-nucleus $i$ is described by $\langle  F_i \rangle,$ which is the overproduction factor of this p-nucleus averaged over the PPLs, i.e.
\begin{equation}
\langle  F_i \rangle = \frac{1}{M_p \, X_{\rm i,ini}} \, \int_{M_{\rm 1}}^{M_{\rm 2}} X_{\rm i}(M_{\rm r}) \, \textrm{d}M_r
\label{eq:Fi}
,\end{equation}
where $M_p = M_2 - M_1$ is the total mass of the PPLs (these layers are delimited by the lower Lagrangian mass coordinate $M_1$ and the upper one, $M_2$), $X_{\rm i,ini}$ the initial abundance of the p-nucleus $i$ and $X_{\rm i}(M_{\rm r})$ the mass fraction of p-nucleus $i$ at mass coordinate $M_{\rm r}$ (after the explosion and beta-decays, if not stated otherwise).


\section{Results}
\label{sect_res}

Fig.~\ref{fig_ab1} shows that the non-rotating model (left panel) only lost $\sim 0.5$~$M_{\odot}$ through winds during the evolution, while the rotating model ejected about $\sim 8$~$M_{\odot}$. 
The main reason for this is that rotation produces larger helium-burning cores, which boosts the stellar luminosity and hence the mass loss. A more luminous star is also more likely to enter the supra-Eddington regime \citep[cf. Sect.~3.2 in][for more details of this model]{choplin18}.

\subsection{s-process nucleosynthesis during hydrostatic burning}
\label{sect_st_mod}

s-process nuclei represent the seed from which p-process elements are made. The s-process has been shown to be affected by rotation during core helium-burning. Consequently, the p-process will also be impacted later in the evolution or during the explosion. 

As shown in Fig.~\ref{fig_ab3}, the s-process becomes more efficient with increasing initial rotation. 
This is due to the stronger operation of the rotational mixing during the core-helium burning phase. It first transports $^{12}$C and $^{16}$O from the He-core to the H-shell, which creates primary $^{14}$N. The $^{14}$N diffuses backward and penetrates the growing convective He-core. This makes primary $^{22}$Ne through the $^{14}$N($\alpha,\gamma$)$^{18}$F($\beta^+$)$^{18}$O($\alpha,\gamma$)$^{22}$Ne chain. The neutron source $^{22}$Ne($\alpha$,$n$)$^{25}$Mg is then boosted, and so is the s-process \citep[e.g.][]{pignatari08, frischknecht12}.

The s-process also becomes more efficient with the $^{17}$O($\alpha$,$\gamma$)$^{21}$Ne reaction rate from \citet[][right panel in Fig.~\ref{fig_ab3}]{taggart19} compared to the rate of \citet[][left panel]{best13}. 
The reason is that the $^{16}$O($n$,$\gamma$)$^{17}$O($\alpha$,$\gamma$)$^{21}$Ne chain is weaker than the competing $^{16}$O($n$,$\gamma$)$^{17}$O($\alpha$,$n$)$^{20}$Ne neutron recycling chain. The latter gives the neutrons back to the medium and hence favours neutron captures by heavy seeds.

In the rotating models computed with the rate of \cite{best13}, rotation boosts the s-process by up to a factor of $10 - 10^3$ for $30 <Z< 60$ with a peak at $Z \sim 45$ (solid lines in Fig.~\ref{fig_ab4}). For a given rotation, the uncertainties associated with the $^{17}$O($\alpha$,$\gamma$)$^{21}$Ne rate change the production by a factor of typically 10, except when fast rotation and heavy elements with $Z>55$ are considered, where the differences reach a factor of about 100 (the solid and dashed red patterns in Fig.~\ref{fig_ab4}).

\begin{figure}
\includegraphics[width=\columnwidth]{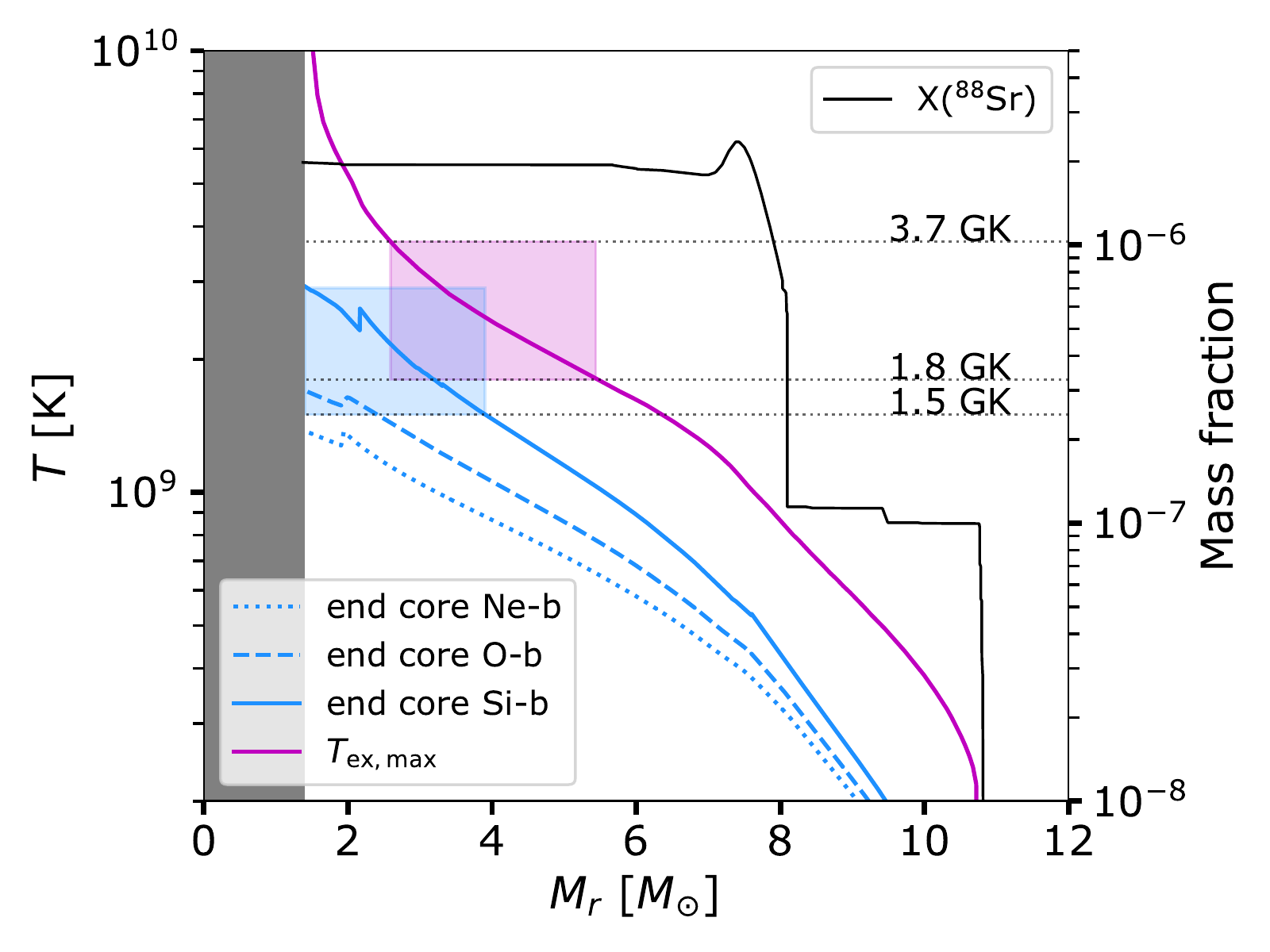}
\caption{
Temperature profiles at different stages of the evolution (blue lines) and maximum temperature reached during the explosion (magenta line) for the M25Zm3V4 model. The abundance profile of the s-nuclide $^{88}$Sr is shown in black with the scale on the right.
The blue and magenta rectangle show where the  p-process can take place (in terms of temperature range and mass coordinate range) during the hydrostatic and explosive burning, respectively (see text for details).
The dark grey area shows the extent of the remnant (corresponding to the size of the $^{56}$Ni core; cf. Sect.~\ref{sect:explomod}). 
The three horizontal dashed lines correspond to temperatures of 1.5, 1.8, and 3.7~GK.
}
\label{fig_tprof}
\end{figure}

\begin{figure}[h!]
\includegraphics[width=0.95\columnwidth]{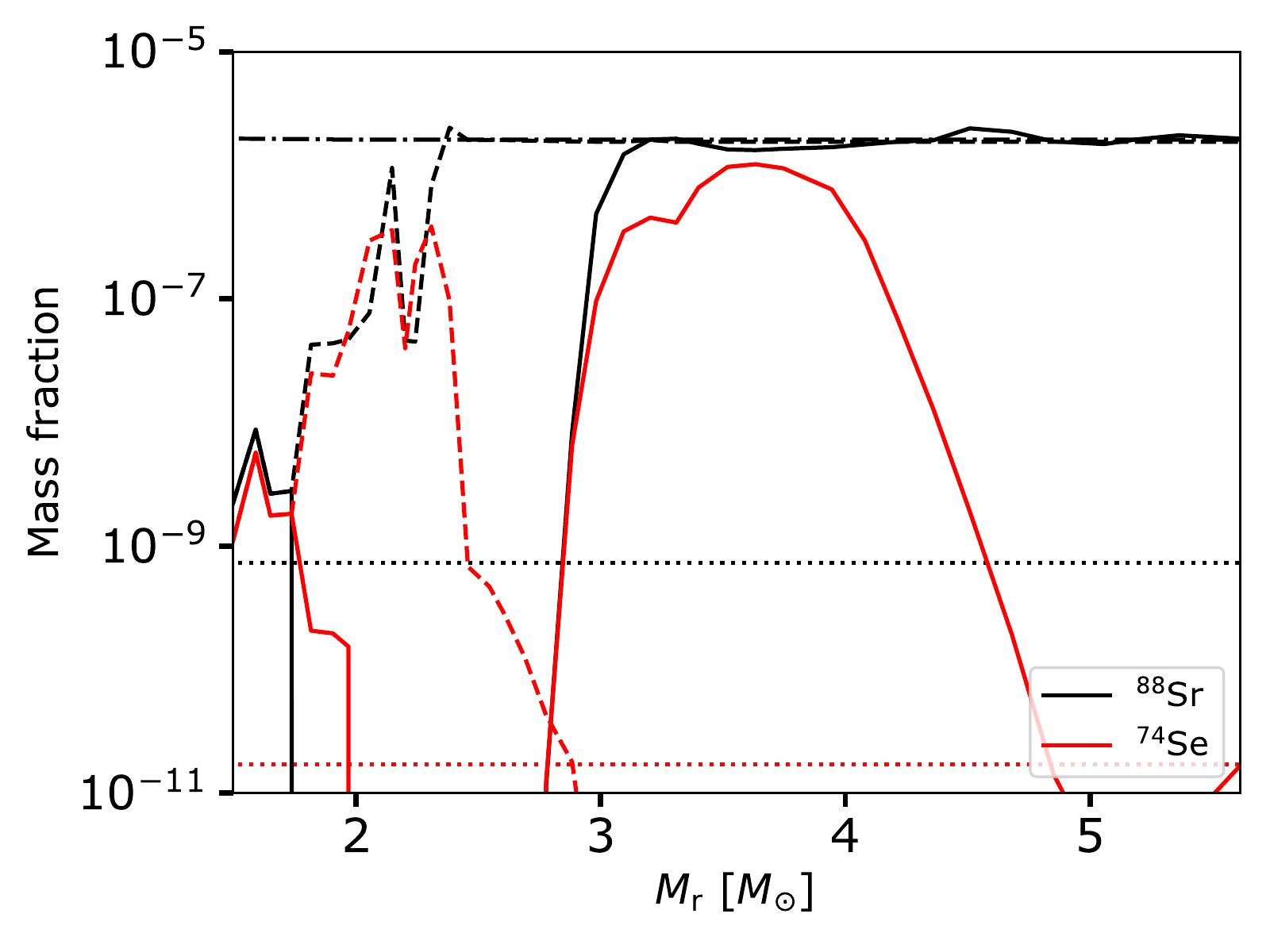}
\caption{
Abundance profiles of the s-nuclide $^{88}$Sr (black) and the p-nuclide $^{74}$Se (red) in the region of the star that is  relevant for the p-process. 
The abundance profiles are shown at four different stages: at the birth of the star (or zero-age main-sequence, dotted lines), at the start of core oxygen burning (dash-dot lines, at this point, the $^{74}$Se abundance is about zero, hence not visible on the plot), at the end of the hydrostatic post-processing calculation (dashed lines) and after the supernova explosion (solid lines). 
}
\label{fig_abprof}
\end{figure}

\begin{figure}[h!]
\includegraphics[width=\columnwidth]{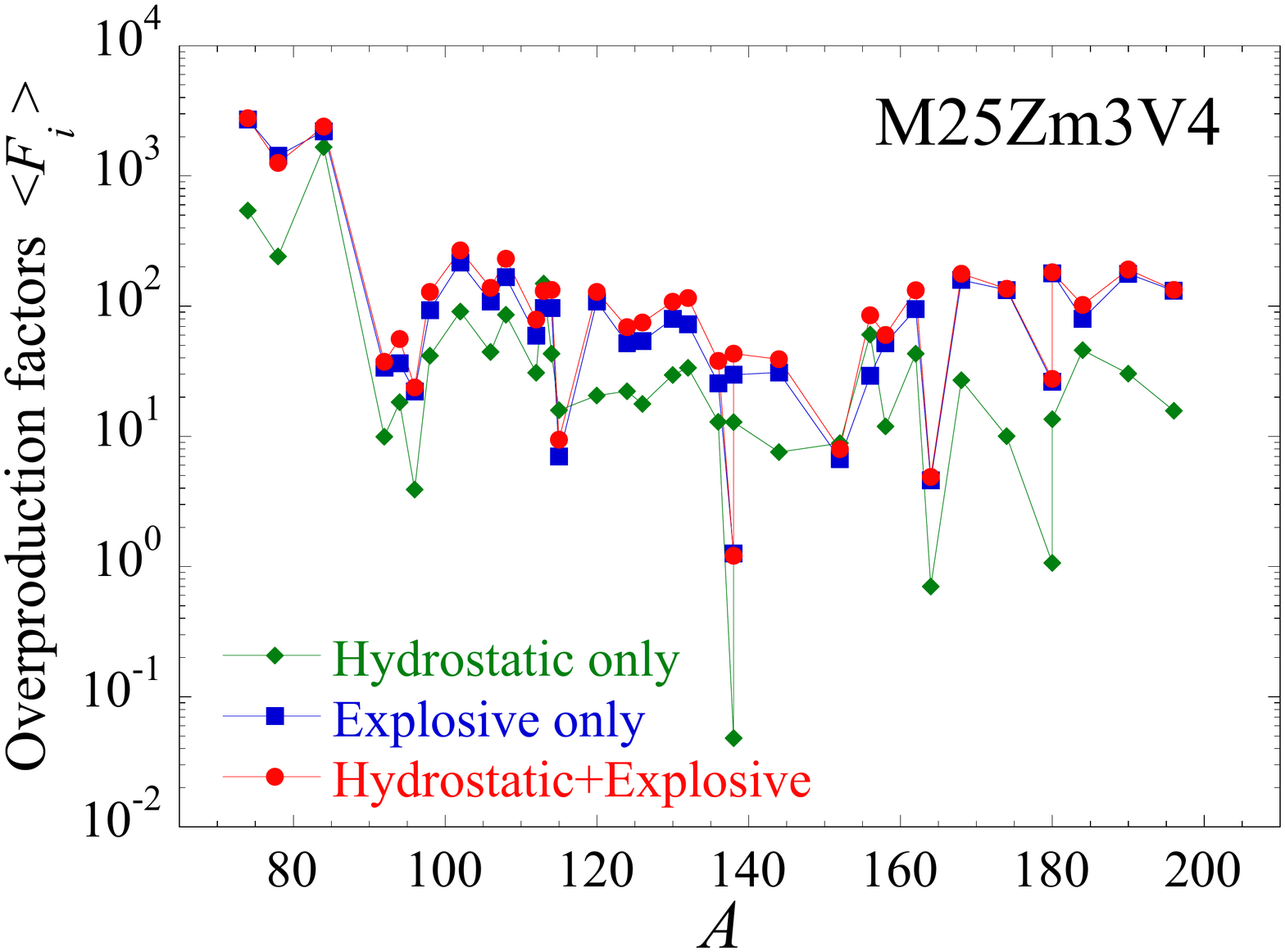}
\caption{Overproduction factors $\langle F_i \rangle$ (Eq.~\ref{eq:Fi}) of p-nuclei mass-averaged over the 4.25~$M_{\odot}$ PPLs (corresponding to the union of the blue and magenta boxes in Fig.~\ref{fig_tprof}) in the M25Zm3V4 model at the end of the hydrostatic burning pre-supernova phase (green diamonds), only due to the explosive burning (blue) or including both the hydrostatic and explosive burning phases (red; see text for more details).}
\label{fig_ppro1}
\end{figure}

\begin{figure*}[h!]
\includegraphics[scale=0.55]{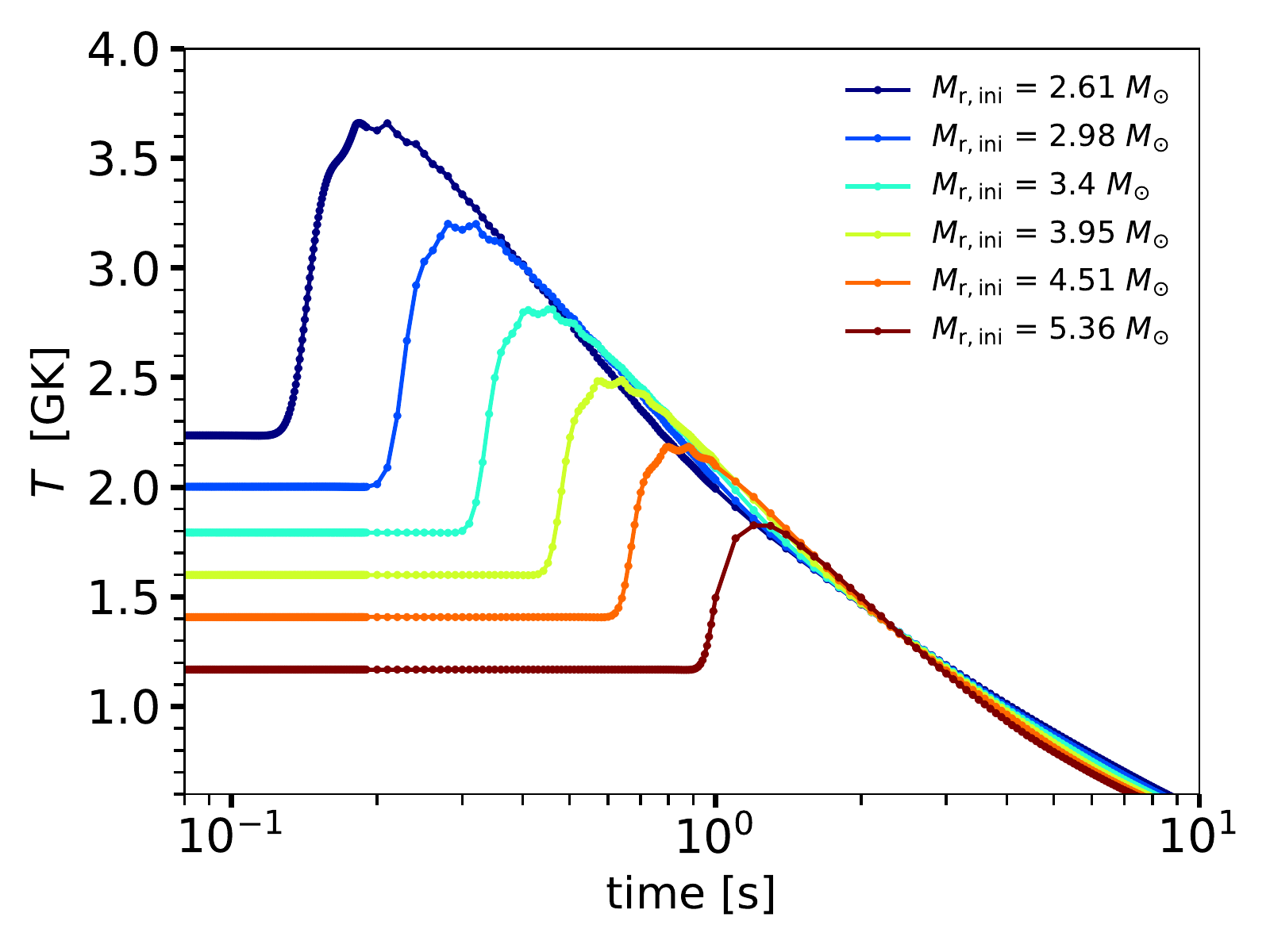}
\includegraphics[scale=0.55]{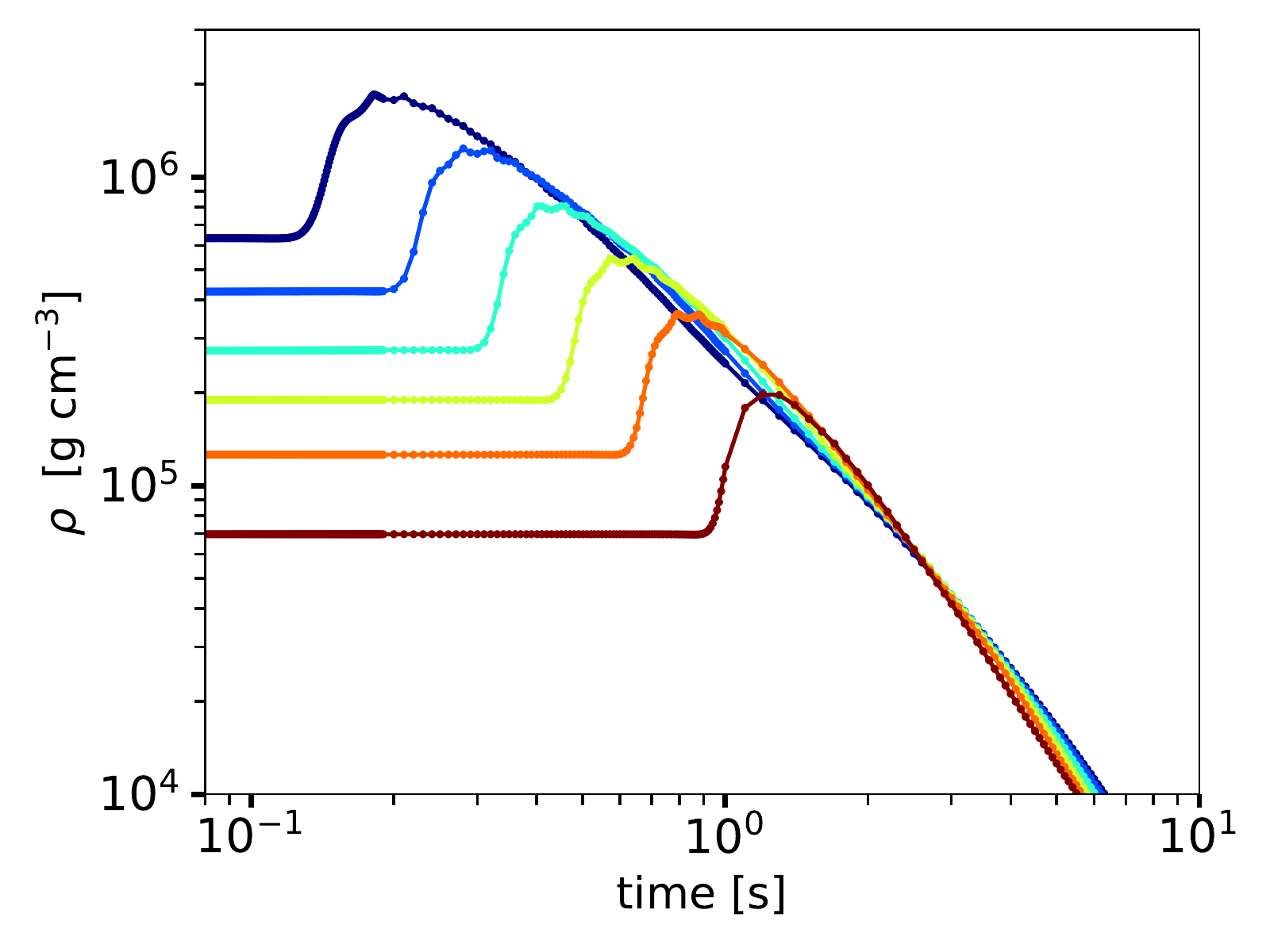}
\caption{
Temperature (left panel) and density (right panel) evolution of several tracer particles coming from the M25Zm3V4 explosion model. The maximum temperature of these particles during the explosion verifies $1.8<T_{\rm ex, max}<3.7$~GK. 
These particles are initially located in the cyan area in Fig.~\ref{fig_ab1} (right panel). 
}
\label{fig_traj}
\end{figure*}

\begin{figure}
\includegraphics[scale=0.45]{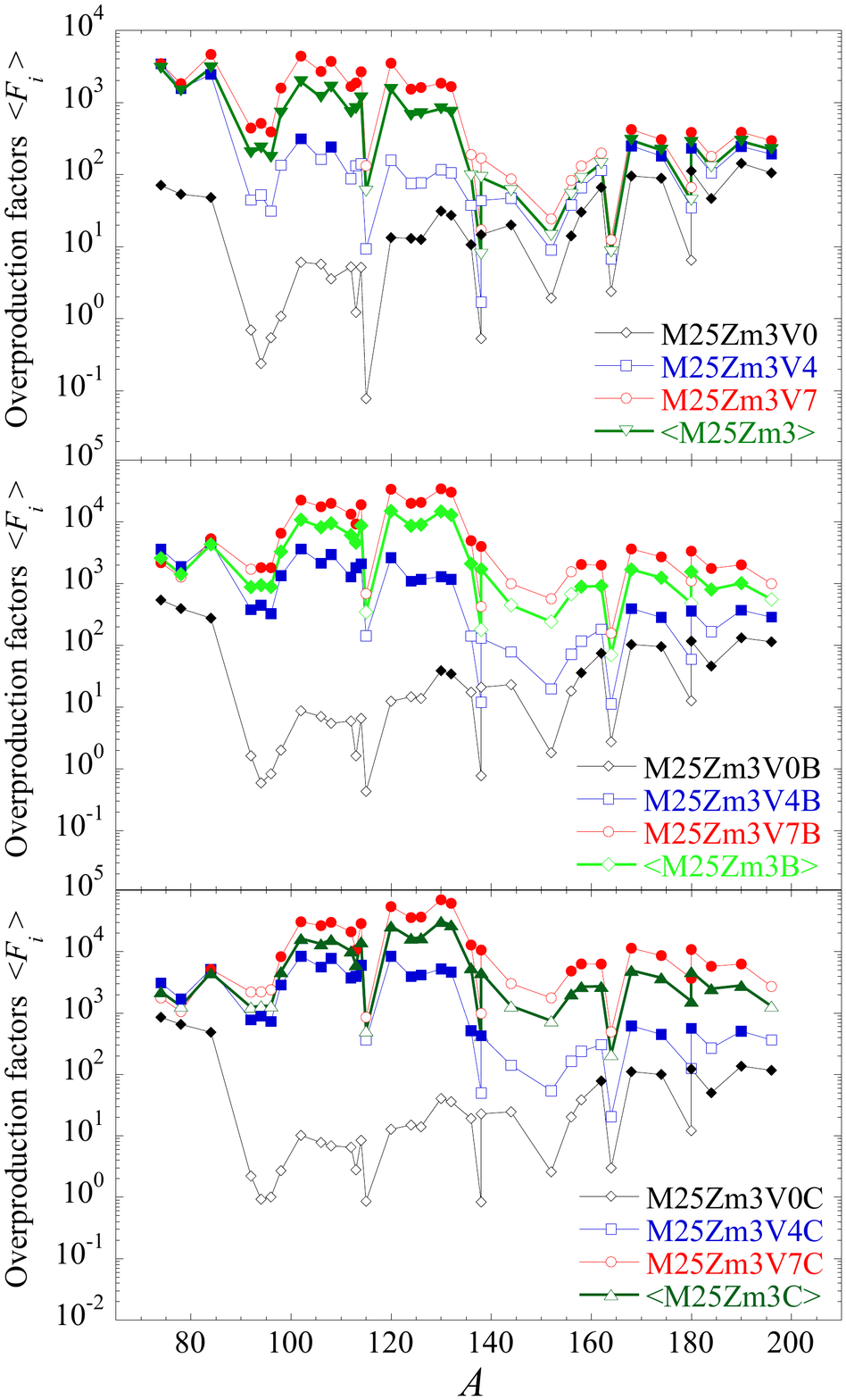}
\caption{Overproduction factors $\langle F_i \rangle$ (Eq.~\ref{eq:Fi}) mass-averaged over all PPLs for all p-nuclides when considering the $^{17}$O($\alpha$,$\gamma$)$^{21}$Ne rate of \citet[][top panel]{best13}, \cite{best13} divided by 10 (middle), and \citet[][bottom panel]{taggart19}. The filled symbols highlight the $n_p$ p-nuclides (Table~\ref{table:2}) that have the highest overabundances, i.e. those with  $\langle F_i \rangle$ higher than the highest value divided by an arbitrary factor of 20. 
The green curves show the velocity-averaged yields over the three 25~$M_{\odot}$ models assuming they  follow the rotational distribution of observed young B stars from \citet[][their Fig. 6]{huang10}. See text for more details. 
}
\label{fig_ppro2}
\end{figure}

\begin{figure}
\includegraphics[width=\columnwidth]{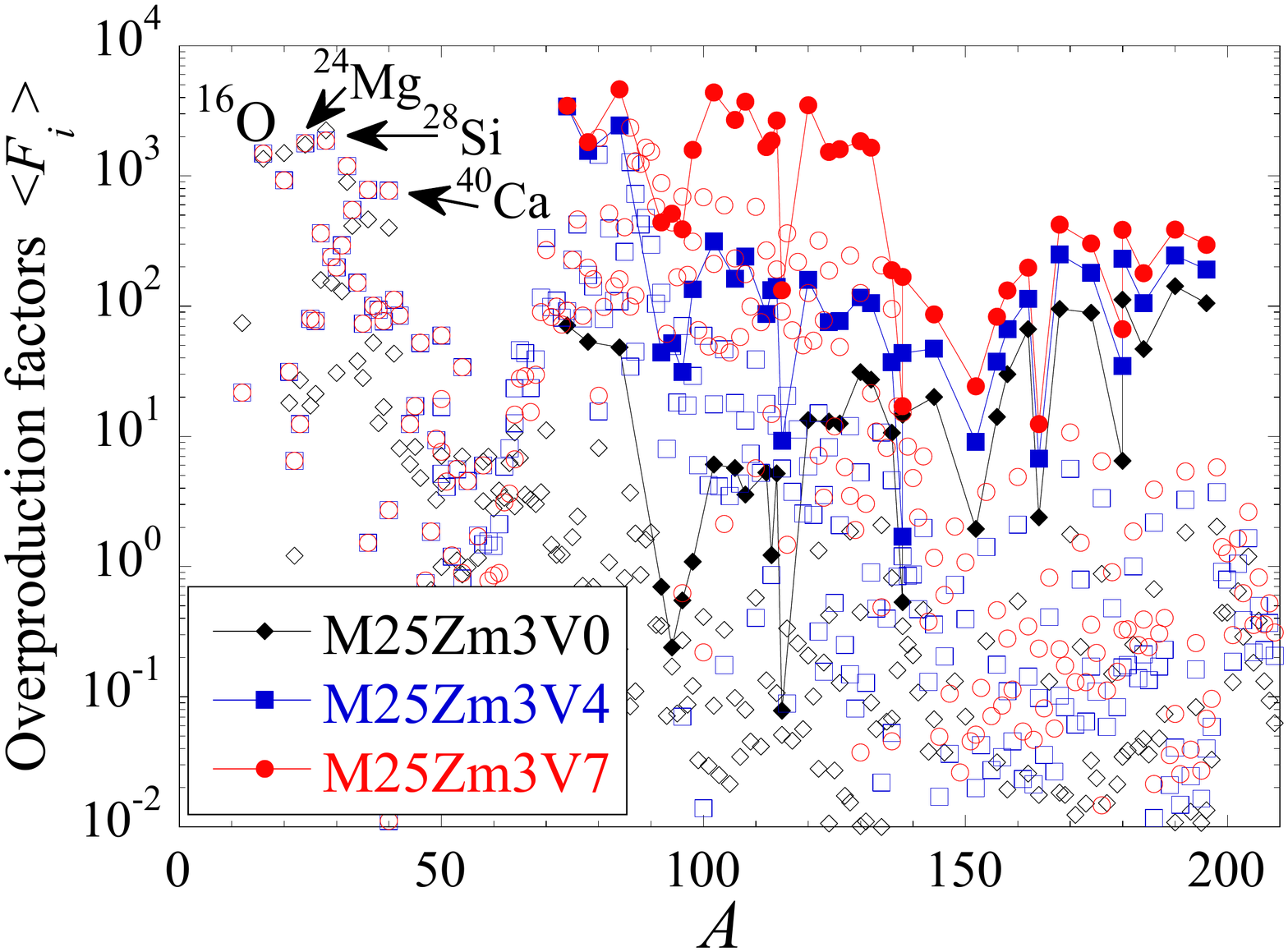}
\caption{Overproduction factors $\langle F_i \rangle$ (Eq.~\ref{eq:Fi}) mass-averaged over all PPLs  for all stable nuclei (open symbols) and p-nuclides (full symbols) for the  25~$M_{\odot}$ models with the $^{17}$O($\alpha$,$\gamma$)$^{21}$Ne rate from \cite{best13} and with $\upsilon_{\rm ini}/ \upsilon_{\rm crit} = 0$ (black diamonds),  $0.4$ (blue squares) and $0.7$ (red circles). The overproduction factors of some $\alpha$-elements, including $^{16}$O, are highlighted.}
\label{fig_ppro3}
\end{figure}

\subsection{p-process nucleosynthesis during hydrostatic burning}
\label{sect_ppro_hydro}

At solar metallicity, some p-nuclides can already form during the ultimate hydrostatic burning stages in the oxygen-burning shell, as initially suggested by \citet{arnould76}. 
Due to the relatively long evolution of the ultimate hydrostatic stages (with respect to the explosion timescale), the p-process during hydrostatic burning already takes place at a minimum temperature of about 1.5~GK \citep{rayet90}. In our M25Zm3V4 model, this temperature was reached at some point during the evolution in the layers with $M_{\rm r} < 3.9 M_{\odot}$. This is illustrated by the light blue area in Fig.~\ref{fig_tprof}.
We evaluated the p-process contribution during hydrostatic burning by post-processing the M25Zm3V4 model. 
We selected the layers that reached a temperature of at least 1~GK before the end of the hydrostatic evolution. These layers were post-processed with our p-process nucleosynthetic code from the onset of core oxygen-burning to the pre-supernova stage. 

Figure~\ref{fig_abprof} shows the abundance profiles of the s-nuclide $^{88}$Sr (black) and p-nuclide $^{74}$Se (red) at different burning stages. 
At the start of core oxygen burning, only $^{88}$Sr is abundant because $^{74}$Se was fully destroyed by the s-process during core helium-burning (dot-dashed black line in Fig.~\ref{fig_abprof}). During the ultimate hydrostatic phases, $^{88}$Sr is partially photodisintegrated at $M_{\rm r} < 2.4$~$M_{\odot}$ (dashed black line) and $^{74}$Se is synthesized around $M_{\rm r} \sim 2.2$~$M_{\odot}$ (dashed red line) through the photodisintegration of s-nuclides.

Figure~\ref{fig_ppro1} shows the results of p-process nucleosynthesis when only the hydrostatic evolution (green), only the explosion (blue, see Sect.~\ref{sect_pexplo} for details; in this case, the abundances of heavy seeds are those at the start of the core oxygen-burning phase), or both the hydrostatic plus explosive burning (red) are considered. 
The contribution stemming from the hydrostatic burning phase is seen to be significant and smaller than the explosive contribution by a factor of $2-5$ for most nuclides. This matter is further processed by the explosive nucleosynthesis. 
Considering only the explosive nucleosynthesis is seen to be an excellent approximation to the total hydrostatic plus explosive processing.
It shows that at least in this model, p-nuclides are mostly synthesized during the explosion. In other conditions, and more particularly, in very massive ($M \ga 100 M_{\odot}$) stars leading to pair creation supernovae, hydrostatic Ne/O burning has been shown to provide a dominant component to the total production of p-nuclei \citep{rayet93}. As proposed by \citet{arnould03}, the p-process may also be found to develop in multi-dimensionally simulated pre-supernova O-rich shells where additional convective mixing is induced by gravity waves \citep{arnett01}. In this case, this extended mixing may also move the p-nuclides that are synthesized before the explosion to external stellar regions, in which they might survive the supernova explosion, in contrast to the pattern found in one-dimensional simulations.

\subsection{p-process nucleosynthesis during the explosion}
\label{sect_pexplo}

We considered explosions of total energy $E_{\rm tot} = 10^{51}$~erg (the impact of the explosion energy is discussed in Sect.~\ref{sect_exp_ene}). The energy was deposited at a mass cut located at the top of the iron ($^{56}$Ni) core. 
During the explosion, the p-process took place in PPLs reaching a peak temperature between $\sim 1.8$ and $\sim 3.5$~GK and corresponding to mass coordinates of $2.59<M_{\rm r}<5.44$~$M_{\odot}$ in our M25Zm3V4 model (magenta area in Fig.~\ref{fig_tprof}). The temperature and density evolution of several PPLs verifying this criteria is shown in Fig.~\ref{fig_traj}.
The relatively high abundance of heavy s-process seeds in this region (represented by the s-nucleus $^{88}$Sr, black line in Fig.~\ref{fig_tprof}) gives rise to an efficient p-process during the explosion. 
This is visible in Fig.~\ref{fig_abprof} for the specific cases of $^{88}$Sr and $^{74}$Se. As other s-nuclides, the pre-supernova $^{88}$Sr abundance (dashed black line) is high above $M_{\rm r} = 2.5$~$M_{\odot}$. This enables an abundant production of p-nuclides such as $^{74}$Se during the explosion (solid red line).
We note that the zone in the star in which the explosive p-process takes place is above the zone in which the hydrostatic p-process takes place (compare the blue and magenta area in Fig.~\ref{fig_tprof}) because the maximum temperature reached during the explosion is higher than in the hydrostatic evolution. The overlap between these regions is about $1.3~M_{\odot}$. This region contains p-nuclides produced  by both hydrostatic and explosive burning. In the inner blue zone (around $M_{\rm r} = 2$~$M_{\odot}$ for instance), the p-nuclides built during hydrostatic burning are all destroyed by photodisintegration due to the high temperatures ($T_{\rm ex, max}>3.7$~GK) encountered during the explosion. 

The s-process efficiency during the evolution increases with initial rotation and with decreasing $^{17}$O($\alpha,\gamma$) rate.
Consequently, the p-process efficiency, which very strongly depends on the initial trans-iron seeds content, also increases with initial rotation and when the $^{17}$O($\alpha,\gamma$) rate is lowered (Fig.~\ref{fig_ppro2}). 
For p-nuclides with $90\la A \la 150$, the overproduction differences between the non-rotating and the fast-rotating models reach about 3 dex at most (Fig.~\ref{fig_ppro2}, black and red curves). This is similar for the differences shown in Figs.~\ref{fig_ab3}-\ref{fig_ab4} (black and red curves) for the s-process. 
For p-nuclides with $A\la 90$ and $A \ga 150$, the differences decrease, as expected from the pre-supernova s-process yields shown in Figs.~\ref{fig_ab3}-\ref{fig_ab4}: they reach about 1 dex at most. 
A lower $^{17}$O($\alpha,\gamma$) rate increases the s-process efficiency (Figs.~\ref{fig_ab3} and \ref{fig_ab4}), and thus also the p-process (Fig.~\ref{fig_ppro2}).
In particular, it leads to the production of a substantial amount of p-nuclides with $A \ga 100$, as expected from the s-process yields shown in Figs.~\ref{fig_ab3}-\ref{fig_ab4}.

Fig.~\ref{fig_ppro3} illustrates the overproduction factors in the rotating 25~$M_{\odot}$ models with $\upsilon_{\rm ini}/ \upsilon_{\rm crit} = 0.4$ and $0.7$, not only for p-nuclei, but also for all stable nuclei. The p-nuclei appear to be rather well co-produced with the $\alpha$-elements such as $^{16}$O, $^{20}$Ne, $^{24}$Mg, or $^{28}$Si. Although the overabundances shown in Fig.~\ref{fig_ppro3} only represent those characterizing the PPLs, the integration over the entire stellar mass gives an overproduction of light p-elements ($A =74-84$ for M25Zm3V4 and $A=74-132$ for M25Zm3V7) similar to those of the light $\alpha$-elements. By contrast, the non-rotating model gives an overproduction factor of about 100 in the PPLs, i.e. about ten times smaller than the factor obtained in such stars for $^{16}$O (black symbols in Fig.~\ref{fig_ppro3}). This difference is due to the low initial metallicity of about $ Z_{\odot} / 10$ adopted here.

Finally, we note that the emitted flux of (anti)neutrinos is not followed in our explosion simulation. For this reason, neutrino-induced nucleosynthesis that can boost the production of some rare p-nuclei, such as the odd-odd $^{138}$La and $^{180}$Ta \citep{Goriely01b,Sieverding18}, is not described in the present work.

\subsection{ $^{92,94}$Mo and $^{96,98}$Ru p-nuclides}
\label{sect_moru}

As shown in Fig.~\ref{fig_ppro2} and found systematically by previous calculations \citep[see e.g.][]{rayet95,arnould03,travaglio18}, the p-nuclides $^{92,94}$Mo and $^{96,98}$Ru are always underproduced by non-rotating models. 
In particular, the overproduction factors of Mo-Ru p-isotopes are smaller by a factor of about 100 than the overproduction factors of the light p-nuclei $^{74}$Se,  $^{78}$Kr, and $^{84}$Sr. 
For rotating models, this deviation is seen to be substantially reduced to a ratio of about 20 and 3 for the models with $\upsilon_{\rm ini}/ \upsilon_{\rm crit} = 0.4$ and 0.7, respectively. In the M25Zm3V7B model with a lower $^{17}$O($\alpha$,$\gamma$)$^{21}$Ne rate (Fig.~\ref{fig_ppro2}, middle panel, red pattern), the Mo-Ru isotopes are overproduced at the same level as the light p-nuclei, but the $100 \la A  \la 132$ p-nuclides are now overproduced with respect to the light nuclides by a factor of about 10. 
The increased production of Mo-Ru p-nuclides is directly linked to the s-enrichment in heavy seeds such as Ba during the evolution of rotating models (cf. Sect.~\ref{sect_st_mod}).

\subsection{Impact of explosion energy on the p-process}
\label{sect_exp_ene}

\begin{figure}[t]
\includegraphics[width=\columnwidth]{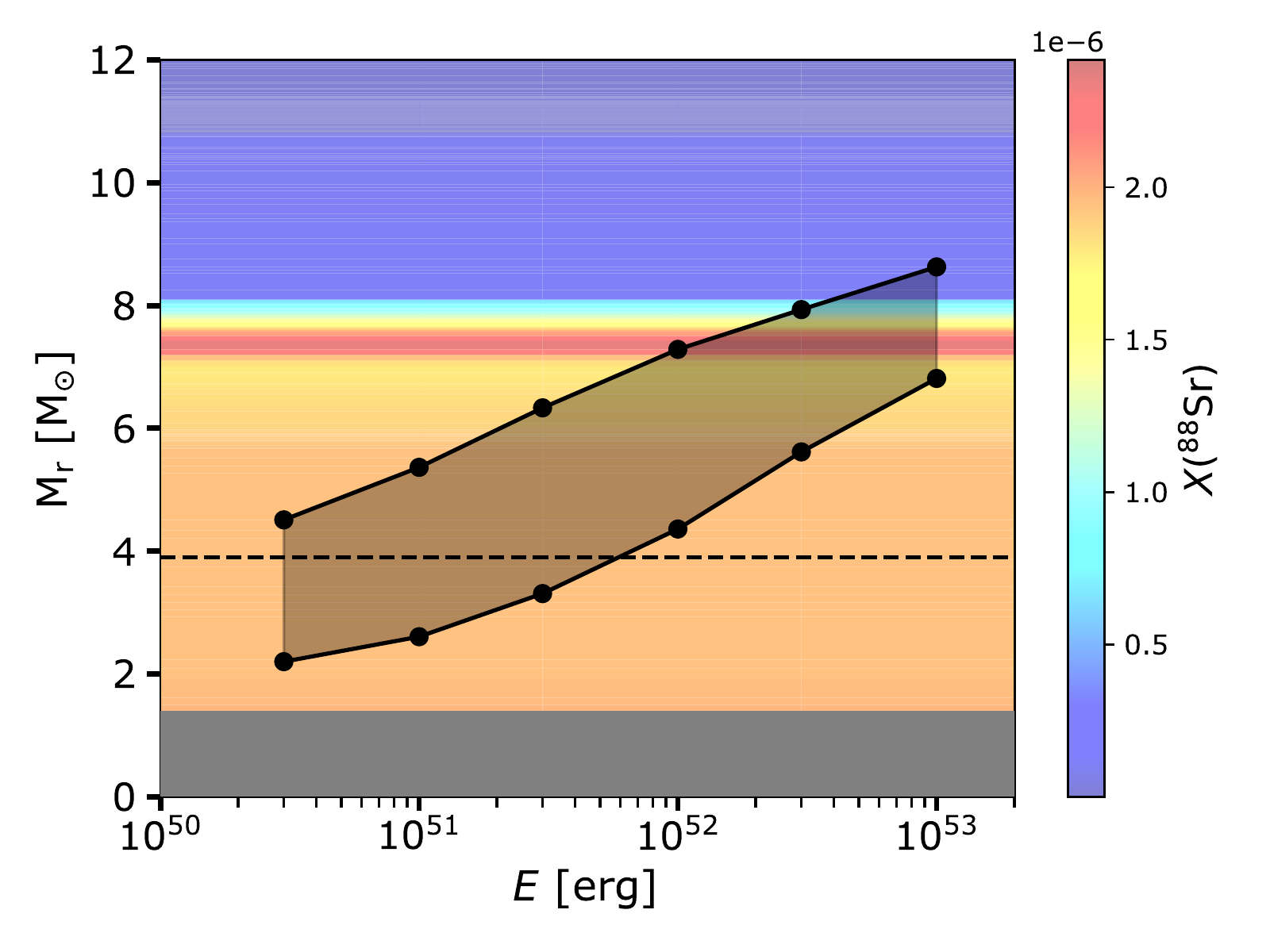}
\caption{
Mass coordinate as a function of explosion energy for the M25Zm3V4 model. 
The color code indicates the mass fraction of the s-nuclide $^{88}$Sr prior to the explosion. The two black lines delimit the zone in which the maximum temperature during the explosion verifies $1.8<T_{\rm ex, max}<3.7$~GK. The dark grey rectangular zone shows the extent of the remnant. The dashed line shows the mass coordinate below which some photodisentegration of trans-iron seeds can have occurred during the last hydrostatic burning stages (zone in which $T>1.5$~GK at some point during hydrostatic burning).
}
\label{fig_expT}
\end{figure}

To obtain an efficient explosive p-process, a region in the star must be enriched in trans-iron seeds and experience temperatures between $1.8$ and $3.7$~GK during the explosion.
In the M25Zm3V4 model, the region enriched in trans-iron seeds extends to about $M_{\rm r} = 8$~$M_{\odot}$ (Figs.~\ref{fig_tprof} and \ref{fig_expT}). Above this, the abundances of trans-iron elements are very similar to their initial abundances. 
The extension of the $1.8<T_{\rm ex, max} <3.7$~GK region is shown in Fig.~\ref{fig_expT} for various explosion energies. 
As expected, this zone shifts upward in mass for more energetic explosions.
Nevertheless, for explosion energies $0.3 \times 10^{51} < E_{\rm tot} < 5 \times 10^{52}$~erg, this zone stays in the stellar region that is enriched in trans-iron seeds.
This implies that an efficient p-process takes place for a whole range of explosion energies in this model. 
Moreover, the yields will be weakly impacted by the explosion energy because the pre-supernova trans-iron seed abundances vary weakly (by less than a factor of $2$) up to $M_{\rm r} = 8$~$M_{\odot}$.

\begin{figure}[t]
\includegraphics[width=\columnwidth]{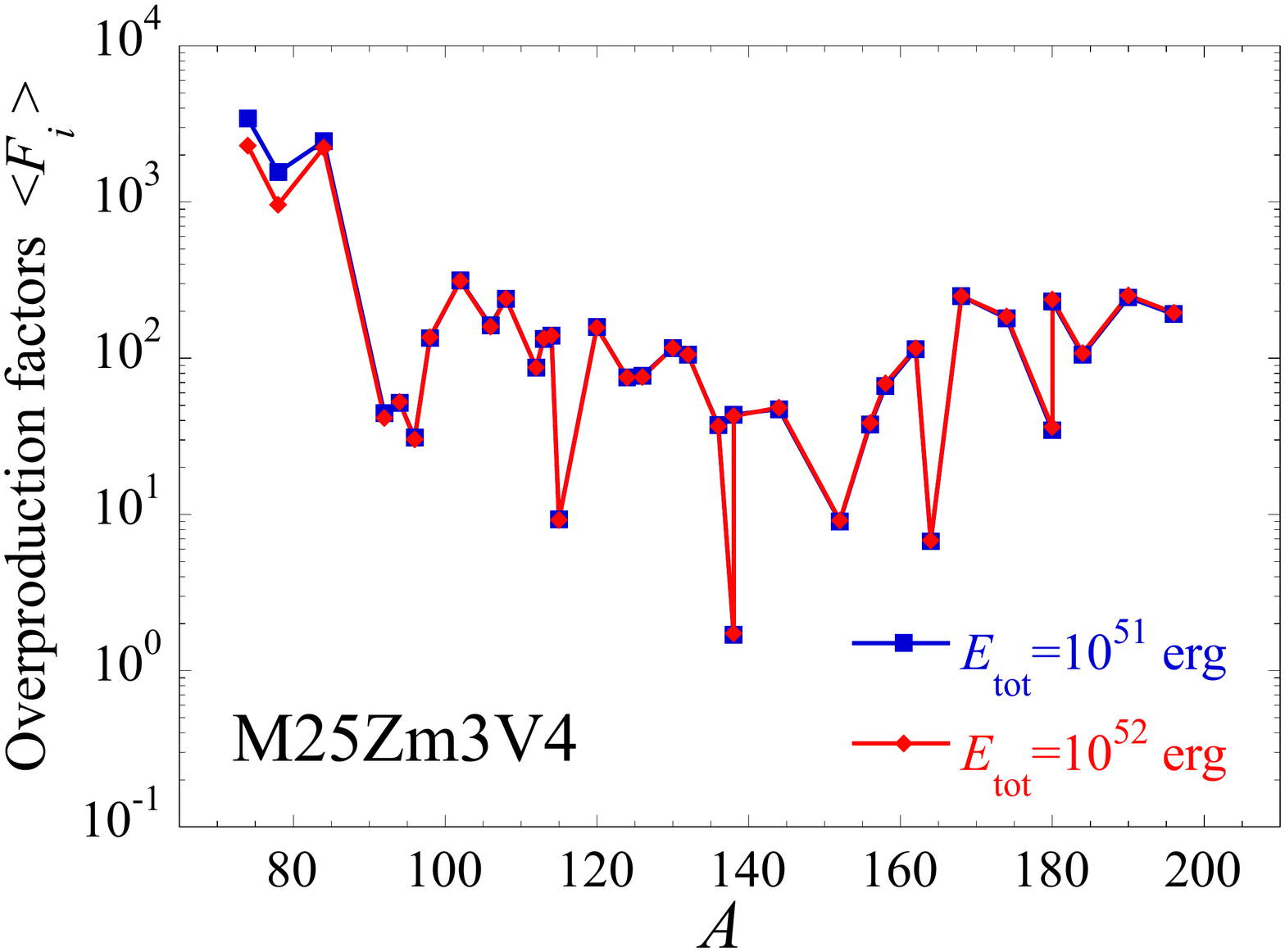}
\caption{Overproduction factors $\langle F_i \rangle$ (Eq.~\ref{eq:Fi}) mass-averaged over all PPLs  for all p-nuclides for the M25Zm3V4 model with two values of the total explosion energy $E_{\rm tot}=10^{51}$ and $10^{52}$~erg. 
}
\label{fig_ppro_etot}
\end{figure}

The resulting abundance distribution of p-nuclides assuming two different values of the explosion energy, namely $E_{\rm tot}=10^{51}$ and $10^{52}$~erg,  are illustrated in Fig.~\ref{fig_ppro_etot}. Both distributions are globally identical, except for the lightest $^{74}$Se and $^{78}$Kr and the $A=180$ Ta and W p-isotopes. As shown in Fig.~\ref{fig_expT}, for $E_{\rm tot}=10^{52}$~erg, the PPLs extend up to the region in which s-process overabundances are highest. For these outer layers, the peak temperature remains relatively low ($T_{\rm ex, max} \simeq 1.8 $~GK), and this consequently essentially boosts the production of the heaviest p-nuclei.

We note that non-spherical explosions can lead to a wide variety of temperature and density conditions compared to the spherical explosions considered in this work. This could ultimately impact the yields of p-nuclides. We plan to investigate the effect of the explosion geometry on the p-process nucleosynthesis in a future work.

\begin{figure}
\includegraphics[scale=0.35, trim=2cm 1cm 0cm 0cm]{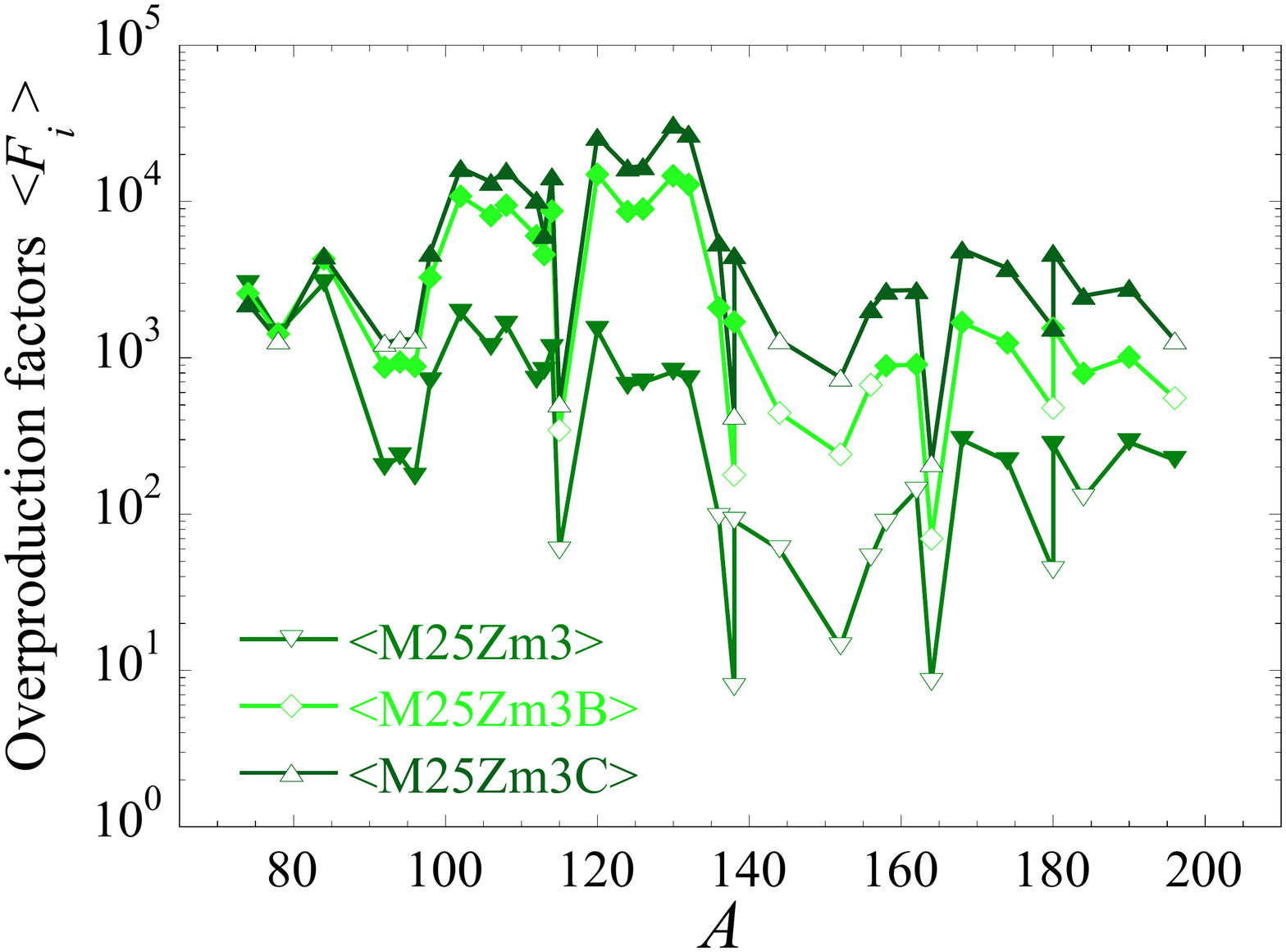}
\caption{Same as Fig.~\ref{fig_ppro2}, where only the overproduction factors for the three  velocity-averaged models are compared.
}
\label{fig_ppro4}
\end{figure}

\subsection{Velocity-averaged yields for a 25~$M_{\odot}$  star}
\label{sect_vaverage}

To make a first rough estimate of the integrated p-nuclide yields of a population of rotating massive stars, we can use the velocity distribution derived by \citet[][their fig. 6]{huang10}. Their velocity distribution was obtained from the observation of 220 young main-sequence B-type stars. We divided their $\upsilon_{\rm ini} / \upsilon_{\rm crit}$ probability density function into three intervals corresponding to our three 25~$M_{\odot}$ models with $\upsilon_{\rm ini} / \upsilon_{\rm crit} = $ 0, 0.4, and 0.7. The first interval extends from $\upsilon_{\rm ini} / \upsilon_{\rm crit}=0$ to $0.2$, the second from 0.2 to 0.55 (midway between 0.4 and 0.7), and the third from 0.55 to 1. 
We then integrated their probability density function over these three intervals to obtain the relative weights of 0.14, 0.45, and 0.41 for the $\upsilon_{\rm ini} / \upsilon_{\rm crit} =$ 0, 0.4, and 0.7 models, respectively. The green curves in Fig.~\ref{fig_ppro2} (also reported in Fig.~\ref{fig_ppro4}) show the resulting overproduction factors when weighting the yields of our three models by the derived coefficients. In each of the three panels of Fig.~\ref{fig_ppro2}, the final yields are mainly influenced by the fast-rotating $\upsilon_{\rm ini} / \upsilon_{\rm crit} =0.7$ model, which indeed has {\it (i)} high overproduction factors and {\it (ii)} a relatively high weight of 0.41. Overall, at this mass and metallicity, the p-nuclide yields of a fast-rotating (typically $\upsilon_{\rm ini} / \upsilon_{\rm crit}=0.7$) massive star are likely well representative of a population of massive stars with various initial rotation rates. 
A more detailed estimation including a finer initial rotational grid, different masses, and metallicities, is deferred to a future work.

\section{Contribution of massive rotating stars to Galactic p-nuclei}
\label{sect_pcontrib}

\begin{table*}
\scriptsize{
\caption{Quantities required to estimate the (O/$p$) ratio corresponding to the oxygen to p-nuclide yield ratio normalized to the solar abundances (Eq.~\ref{eq:op}). $M_{\rm CO}$ (column 5) is the mass of the carbon-oxygen core (or helium-free core), $M_p$ (column 6) is the mass of the layers undergoing p-process nucleosynthesis during the explosion, $\mathcal{Y}_{\rm O}$ (column 7) is the yield (total ejected mass) of oxygen, $F_0$ (column 8) is the average overproduction factor (Eq.~\ref{eq:f0}), $F_1$ (column 9) is like $F_0$ , but the sum runs only over the $n_p$ (column 10) most abundant p-nuclides, (O/$p$)$_0$ (column 11) and (O/$p$)$_1$ (column 12) are computed with $F_0$ and $F_1$, respectively. The $<$M25Zm3$>$, $<$M25Zm3B$>,$ and $<$M25Zm3C$>$ rows correspond to the velocity-averaged yield of the three models above (cf. Sect.~\ref{sect_vaverage} for details). }
\label{table:2}
}
\begin{center}
\resizebox{17cm}{!} {
\begin{tabular}{lccccccccccc} 
\hline
Model label  & $M_{\rm ini}$ &   $\upsilon_{\rm ini}/ \upsilon_{\rm crit} $    & $^{17}$O($\alpha$,$\gamma$)$^{21}$Ne &  $M_{\rm CO}$ &  $M_p$  & $\mathcal{Y}_{O}$  &  $F_0$  & $F_1$  &   $n_p$  &  (O/$p$)$_0$ &  (O/$p$)$_1$    \\
          & [$M_{\odot}$]  &     &       &  [$M_{\odot}$]       &[$M_{\odot}$]&[$M_{\odot}$]  &  &  &   \\
\hline
M25Zm3V0   & 25 & 0 & \cite{best13} &         5.90                 &    2.37   &    2.75     &   30.2   & 50.8   &  20 &      62.0      &  37.1    \\
M25Zm3V4   & 25 & 0.4 & \cite{best13} &          7.61             &    2.93    &    4.56         &  312  &  908   & 10 &   7.30      &   2.50       \\
M25Zm3V7   & 25 & 0.7 & \cite{best13} &         7.56            &    2.93      &    5.30      & 1229  &   1814  & 23 &      2.13      &    1.39   \\
$<$M25Zm3$>$ & $-$ & $-$ & \cite{best13}  &        7.35          &   2.93   &     4.61          &  649    &  952  & 23  &    3.55      &   2.41  \\
\hline
M25Zm3V0B & 25 & 0 & \cite{best13} / 10   &         5.88          &   2.37   &   2.91        &   62.3   &  154  &  13 &   30.0       & 11.8    \\
M25Zm3V4B & 25 & 0.4 & \cite{best13} / 10   &       7.64            &   2.93   &    5.23       &   1047   &    1544 & 23  & 2.49         &   1.69  \\
M25Zm3V7B & 25 & 0.7 & \cite{best13} / 10   &       7.62           &   2.93   &     5.27    &    8373  &  11346  &  25 & 0.31         &   0.23  \\
$<$M25Zm3B$>$ & $-$ & $-$ &\cite{best13} / 10 &       7.39           &    2.93  &  4.92   &   3914   &  4963  & 27  &  0.63         &  0.49   \\
\hline
M25Zm3V0C & 25 & 0 & \cite{taggart19}   &         5.88          &  2.37    &      2.91     &   86.3   &  271  & 10  & 21.4         & 6.68    \\
M25Zm3V4C & 25 & 0.4 & \cite{taggart19}   &       7.65            & 2.93     &      5.21     &  2353    &  3346  & 24  & 1.10         &  0.78   \\
M25Zm3V7C & 25 & 0.7 & \cite{taggart19}   &       7.57            &   2.93   &     5.28    &  15008    &  21084  &  24 &  0.18        &   0.12  \\
$<$M25Zm3C$>$ & $-$ & $-$ &\cite{taggart19} &       7.37           &  2.93    &   4.91  &  7229    & 9734   &  25 & 0.34         &  0.25   \\
\hline
\end{tabular}
}
\end{center}
\end{table*}

Since oxygen is known to be essentially produced by CCSNe, the contribution of massive stars to the Galactic p-nuclei enrichment can be estimated through the oxygen to p-nuclide yield ratio normalized to the solar abundances, (O/$p$), as defined by \citet{rayet95}. The calculation of this quantity requires the estimate of the average overproduction factor $F_0$ over the 35 p-nuclides given by
\begin{equation}
F_0 = \sum_i \, \langle  F_i \rangle / 35
\label{eq:f0}
,\end{equation}
where $\langle  F_i \rangle$ is the overproduction factor of p-nuclide $i$ averaged over the PPLs (Eq.~\ref{eq:Fi}).
We note that we considered  only the explosive contribution of the p-process to estimate the (O/$p$) ratios (which is a good approximation, as shown in Fig.~\ref{fig_ppro1}). Therefore, the PPLs mass $M_p$ used in Eq.~\ref{eq:Fi} and given in Table~\ref{table:2} corresponds to the zone in the star that experiences p-process  only during the explosion (magenta zone in Fig.~\ref{fig_tprof}). 

The net yield $y_{\rm i}$ of a p-nuclide $i$ is defined as the difference between the mass of a p-nuclide $i$ returned by the star to the interstellar medium and the mass of that same p-nuclide engulfed at the birth of the star. In our case, it can be simplified as follows:
\begin{equation}
y_{\rm i} = X_{\rm i,ini} \, F_0 \, M_p - X_{\rm i,ini} \, M_{\rm CO}
\label{eq:yi}
,\end{equation}
where $M_{\rm CO}$ is the carbon-oxygen core mass, which also corresponds to the helium-free core (Table~\ref{table:2}). 
As in \citet{rayet95}, Eq.~\ref{eq:yi} assumes that the p-nuclei initially present in the helium-burning core have been destroyed by the s-process and that their production in the PPLs can be averaged by the same overproduction factor $F_0$. 

To derive the (O/$p$) ratio, a similar estimate of the net oxygen yield can be derived by
\begin{equation}
y_{\rm O} = \mathcal{Y}_{\rm O} - X_{\rm O,ini} \, M_{\rm ini}
\label{eq:yO}
,\end{equation}
where $\mathcal{Y}_{\rm O}$ is the absolute oxygen yield (i.e. the total mass  of oxygen ejected through stellar winds and supernova, as reported in Table~\ref{table:2}), $X_{\rm O,ini}$ is the initial oxygen mass fraction, and $M_{\rm ini}$ is the initial mass. 
The supernova ejecta enrich the interstellar medium with oxygen and p-nucleus $i$ in solar proportion if
\begin{equation}
\frac{y_{\rm O}}{y_{\rm i}} = \frac{X_{\rm O,\odot}}{X_{\rm i,\odot}}
\label{eq:yoyi}
,\end{equation}
where $X_{\rm O,\odot}$ and $X_{\rm i,\odot}$ are the solar mass fraction of oxygen and p-nuclide $i$, respectively.
Combining Eq.~\ref{eq:yoyi} with Eqs.~\ref{eq:yi} and \ref{eq:yO} gives the (O/$p$)$_0$ ratio for a given averaged p-enrichment, given by $F_0$ as
\begin{equation}
(\textrm{O}/p)_0 \equiv \frac{y_{\rm O} }{y_{\rm i} }   \frac{X_{\rm i,\odot}}{X_{\rm O,\odot}} = \frac{\mathcal{Y}_{\rm O}/X_{\rm O,\odot} - f_Z M_{\rm ini}}{f_Z F_0 M_p - f_Z M_{\rm CO}}
\label{eq:op}
,\end{equation}
where $f_Z= Z/Z_{\odot} = 0.001 / 0.014 = 0.071$ in our case. 
We note that considering a simple average overproduction factor $F_0$ may not be optimal in our case because of the relatively large dispersion of the overproduction factors $\langle F_i \rangle$ in some of our models (Fig.~\ref{fig_ppro2}).
Therefore we also define an averaged overproduction factor $F_1$ , which in contrast to $F_0$ only considers the most abundant p-nuclei. This new factor $F_1$ is defined by an equation similar to Eq.~\ref{eq:f0}, but the sum runs only over the $n_p$ p-nuclides (given in Table~\ref{table:2}), the mass fraction of which is greater than $X_{\rm max}/20,$ where $X_{\rm max}$ is the mass fraction of the most abundant p-nucleus. 
These $n_p$ p-nuclides are highlighted with filled symbols in Fig.~\ref{fig_ppro2}.
The factor of 20 considered here is purely arbitrary and aims at considering the various uncertainties associated with both the astrophysical modelling (including those affecting the s-process in massive stars) and the nuclear ingredients \citep[see in particular Fig.~35 of ][]{arnould03}. Therefore, we define two ratios (O/$p$)$_0$ and  (O/$p$)$_1$ depending on whether the (O/$p$) ratio in Eq.~\ref{eq:op} is calculated with $F_0$ or $F_1$ (Table~\ref{table:2}).  We also computed the (O/$p$) ratios for the velocity-averaged models, as reported in the fourth row of each series of models in Table~\ref{table:2}. 

If (O/$p$)~$= 1$, the p-nuclides are produced in solar proportion with oxygen on average. However, if our models predict (O/$p$)~$>1$, p-nuclides are predicted to be underproduced compared to oxygen. This would imply that our models alone cannot fully account for the p-nuclide enrichment in the Solar System, and, more generally, in the Galaxy. As shown in Table~\ref{table:2},   $0.17< \, $(O/$p$)$_0 < 61.3$ and $0.12 < \, $(O/$p$)$_1 < 36.9 $. The (O/$p$) ratios rapidly drop with increasing rotation and with decreasing $^{17}$O($\alpha,\gamma$) rate, meaning that more p-nuclei are produced compared to oxygen if massive stars are rotating or if the $^{17}$O($\alpha,\gamma$) rate is lower (as already discussed in Sect.~\ref{sect_res}). 

\cite{rayet95} considered solar metallicity non-rotating massive stars with initial masses ranging between 13 and 25~$M_{\odot}$. They found $1.8 < \, $(O/$p$)$_{0} < 8.4$ and a ratio of 4.2 when they integrated over the initial mass function from \cite{kroupa93}. 
In our case, the (O/$p$)$_{0}$ ratios for non-rotating models are significantly higher, with values ranging between 6.7 and 61.3 (Table~\ref{table:2}).
Nevertheless, their results are compatible with ours in view of the different metallicities considered. Oxygen is a primary product, in contrast to s- and p-nuclei, which are secondary in non-rotating models and are hence metallicity dependent. \cite{rayet95} concluded that about one quarter of the solar system p-nuclei could be attributed to supernovae from massive stars (assuming that the entire solar oxygen is coming from such events). 
Similarly, on the basis of a Galactic chemical evolution model, \cite{travaglio18} estimated that CCSNe from populations of non-rotating massive stars with different initial masses and metallicities could contribute no more than 10\% to the Galactic p-enrichment, with only a few exceptions (e.g. the light p-nuclides $^{74}$Se, $^{78}$Kr, and $^{84}$Sr). The contribution stemming from models with sub-solar metallicities was also found to be negligible due to the secondary nature of the s- and p-processes.

Our results with (O/$p$) ratios close to or even below one (Table~\ref{table:2}) suggest that sub-solar-metallicity rotating massive stars can co-produce or even overproduce p-nuclei with respect to oxygen. The impact of rotation on the s-process was found to be so large in stars with a metallicity $Z\simeq 0.001$ \citep{frischknecht16, choplin18, choplin20} that it gives rise to large overproduction factors of p-nuclei (see Table~\ref{table:2} and Fig.~\ref{fig_ppro2}) that can compensate for their sub-solar metallicity. 
Table~5 in \cite{frischknecht16} shows that the overproduction factors of light s-nuclides for their solar metallicity models are $\sim 50$ at most and that rotation increases them by no more than a factor of about 2. 
The overproduction factors of p-nuclides for these models therefore is not significantly affected by rotation and is characterized by averaged values of $F_0\simeq 100$ \citep[see e.g. Table 3 in][]{rayet95}.
The averaged overproduction factors of our models with respect to the solar abundances $F_{0, \odot}$ and $F_{1, \odot}$ 
can also be estimated from our Table~\ref{table:2} by simply dividing the $F_0$ and $F_1$ factors by the $Z_{\odot}/Z = 14$ ratio. Considering the velocity-averaged yields of p-nuclides, overproduction factors of $F_{0, \odot} \simeq 50$ and $F_{1, \odot} \simeq 70$ are found when considering the rates of \cite{best13}. When using the rate of \cite{best13} divided by 10, we find $F_{0, \odot} \simeq 280$ and $F_{1, \odot} \simeq 350$. Finally, $F_{0, \odot} \simeq 520$ and $F_{1, \odot} \simeq 700$ if using the rate of \cite{taggart19}. 
These value are similar to or higher than those expected from solar metallicity non-rotating and rotating massive stars. 
On this basis, sub-solar-metallicity rotating stars may even be dominant contributors to the Galactic enrichment of p-nuclei (and s-nuclei) compared to solar metallicity (rotating) stars. 
Detailed Galactic evolution models are needed to quantify their global contribution. Such a quantitative study is postponed to a future work.


\section{Summary and conclusions}
\label{sect_concl}

We studied the p-process in rotating massive stars during their ultimate hydrostatic burning stages and during their explosions. 
These stars can experience an enhanced s-process during their evolution due to the effect of rotational mixing. 
Consequently, they are enriched in trans-iron seeds in the last evolutionary stages and during the explosion. 
Because of the higher abundance in heavy seeds, they are expected to experience an efficient p-process nucleosynthesis during their explosion and possibly during their last hydrostatic burning stages. 

We computed 25~$M_{\odot}$ stellar models at a metallicity of $Z=10^{-3}$ with different initial rotation velocities and $^{17}$O($\alpha$,$\gamma$)$^{21}$Ne rates. 
We found that most of the p-nuclides are synthesized during the explosion. 
The impact of rotation on the p-process  follows the effect of rotation on the s-process. 
From no to fast rotation, both the s- and p-process efficiencies are boosted by about $3-4$~dex for nuclides with $A<140$. The impact remains small for nuclides with $A \geq 140$, unless both fast rotation and a lower $^{17}$O($\alpha$,$\gamma$)$^{21}$Ne reaction rate are considered. In this case, a significant number of heavy p-nuclides (with $A \geq 140$) is produced. 
The impact of the explosion energy on the p-process yields remains weak because a different explosion energy just shifts the zone in the star in which the p-process takes place. Because this zone is similarly enriched in s-nuclides (built during the previous evolutionary stages), the p-process yields are weakly impacted.

By considering a population of solar-metallicity non-rotating massive stars, \cite{rayet95} and  \cite{travaglio18} concluded that CCSNe from populations of non-rotating massive stars could contribute no more than $10 - 25$~\% to the Galactic p-enrichment, and as in previous works, they reported that sub-solar-metallicity stars play a minor role. However, their studies only considered non-rotating models and may consequently have underestimated the role of CCSNe in the Galactic enrichment of p-nuclei. Our present results suggest that rotating massive stars with sub-solar metallicity may substantially contribute to the Galactic p-enrichment, and that the global contribution stemming from CCSNe needs to be revisited. In particular, rotating massive stars with sub-solar metallicity may be the dominant contributors compared to solar metallicity massive stars because the effect of rotational mixing at sub-solar metallicity is stronger. However, before drawing quantitative conclusions, more detailed studies are required that use a larger grid of masses, metallicities, and rotation velocities, supplemented by Galactic chemical evolution simulations.

The p-process efficiency in rotating massive stars also remains very strongly impacted by the uncertainty associated with the $^{17}$O($\alpha$,$\gamma$)$^{21}$Ne reaction rate and its effects on the s-process seed distribution. An accurate determination of this rate will greatly help constrain the contribution of massive rotating stars to the Galactic content not only in s-nuclei, but also in p-nuclei. 

As mentioned at the end of Sect.~\ref{sect_exp_ene}, the p-process nucleosynthesis in self-consistent multi-dimensional explosions from rotating progenitors also remains to be assessed. 
These models are thought to experience r-process nucleosynthesis \citep{winteler12,nishimura15,mosta18,reichert21}. 
Our simulations suggest that exploding rotating massive stars can be rich nucleosynthesis astrophysical sites for the production of elements heavier than iron, in particular by enriching the interstellar medium in (1) s-process material  from their outer He-rich layers, (2) p-process material from deep O-rich layers, and (3) some r-process material from the innermost ejecta.

Finally, our study once again highlights the important impact of mixing mechanisms in deep stellar interiors on the nucleosynthesis in general, and of the p-process in particular. 
The only way to substantially improve the current situation in this respect is to master multi-dimensional star simulations in future developments.

\section*{Acknowledgments}
We wish to thank M. Arnould for his careful reading and valuable remarks. 
This work was supported by the Fonds de la Recherche Scientifique-FNRS under Grant No IISN 4.4502.19. 
S.G. is FRS-F.N.R.S. research associate. 
RH acknowledges support from the World Premier International Research Centre Initiative (WPI
Initiative, MEXT, Japan), STFC UK, the European Union’s Horizon 2020 research and innovation programme under grant agreement No 101008324 (ChETEC-INFRA) and the IReNA AccelNet Network of Networks, supported by the National Science Foundation under Grant No. OISE-1927130. RH also acknowledges support from the ChETEC COST Action (CA16117), supported by COST (European Cooperation in Science and Technology).
GM has received funding from the European Research Council (ERC) under the European Union's Horizon 2020 research and innovation programme (grant agreement No 833925, project STAREX).

\bibliographystyle{aa}
\bibliography{astro}

\end{document}